\documentclass{article}

\usepackage{authblk}
\usepackage{amsmath}
\usepackage{amsfonts}
\usepackage{graphicx,subcaption}
\usepackage{url}

\usepackage{graphicx}
\usepackage{dcolumn}
\usepackage{bm}
\usepackage{color}

\usepackage[utf8]{inputenc}
\usepackage[T1]{fontenc}
\usepackage{mathptmx}
\usepackage{amsthm}
\usepackage{subcaption}
\usepackage{hyperref}

\title{A generic model for pandemics in networks of communities and the role of vaccination}

\author[1]{Chris G. Antonopoulos}

\author[2]{Mohammad H. Akrami}

\author[3,4]{Vasileios Basios}

\author[5]{Anouchah Latifi}

\affil[1]{Department of Mathematical Sciences, University of Essex, UK\thanks{E-mail: canton@essex.ac.uk}}

\affil[2]{Department of Mathematics, Yazd University, Iran\thanks{E-mail: akrami@yazd.ac.ir}}

\affil[3]{Service de Physique des Syst\`emes Complexes et M\'ecanique Statistique, Universit\'e Libre de Bruxelles, Belgium}

\affil[4]{Interdisciplinary Center for Nonlinear Phenomena and Complex Systems (CeNoLi), Belgium\thanks{E-mail: vbasios@ulb.ac.be}}

\affil[5]{Department of Mechanics, Qom University of Technology, Iran\thanks{E-mail: latifi@qut.ac.ir}}

\date{\today}

\begin{document}

\maketitle
\begin{abstract}
\noindent The slogan ``nobody is safe until everybody is safe'' is a dictum to raise awareness that in an interconnected world, pandemics such as COVID-19, require a global approach. Motivated by the ongoing COVID-19 pandemic, we model here the spread of a virus in interconnected communities and explore different vaccination scenarios, assuming that the efficacy of the vaccination wanes over time. We start with susceptible populations and consider a susceptible-vaccinated-infected-recovered model with unvaccinated (``Bronze''), moderately vaccinated (``Silver'') and very well vaccinated (``Gold'') communities, connected through different types of networks via a diffusive linear coupling for local spreading. We show that when considering interactions in ``Bronze''-``Gold'' and ``Bronze''-``Silver'' communities, the ``Bronze'' community is driving an increase in infections in the ``Silver'' and ``Gold'' communities. This shows a detrimental, unidirectional effect of non-vaccinated to vaccinated communities. Regarding the interactions between ``Gold'', ``Silver'' and ``Bronze'' communities in a network, we find that two factors play central role: the coupling strength in the dynamics and network density. When considering the spread of a virus in Barab\'asi-Albert networks, infections in ``Silver'' and ``Gold'' communities are lower than in ``Bronze'' communities. We find that the ``Gold'' communities are the best in keeping their infection levels low. However, a small number of ``Bronze'' communities are enough to give rise to an increase in infections in moderately and well-vaccinated communities. When studying the spread of a virus in a dense Erd\H{o}s-R\'enyi, and sparse Watts-Strogatz and Barab\'asi-Albert networks, the communities reach the disease-free state in the dense Erd\H{o}s-R\'enyi networks, but not in the sparse Watts-Strogatz and Barab\'asi-Albert networks. However, we also find that if all these networks are dense enough, all types of communities reach the disease-free state. We conclude that the presence of a few unvaccinated or partially vaccinated communities in a network, can increase significantly the rate of infected population in other communities. This reveals the necessity of a global effort to facilitate access to vaccines for all communities.
\end{abstract}

\vskip 1cm

\noindent {\bf Keywords:} COVID-19, Pandemics, SIR model, SVIR model, Mathematical modelling, Epidemiology, Vaccination, Networks, Complex systems
\vskip 1cm

\noindent\textbf{Motivated by the ongoing COVID-19 pandemic and vaccination programs implemented worldwide, we model here mathematically the spread of a virus in interconnected communities, which can be for example countries, and study different vaccination scenarios, assuming that the efficacy of vaccination wanes over time. These scenarios result in different infection outcomes, depending on how the communities are interconnected and how strongly interconnected they are. We consider a susceptible-vaccinated-infected-recovered model and initially susceptible populations in communities which can be unvaccinated (``Bronze''), moderately vaccinated (``Silver'') or very well vaccinated (``Gold''). We connect them through different types of networks, i.e., through Erd\H{o}s-R\'enyi, Watts-Strogatz and Barab\'asi-Albert networks, via a diffusive linear coupling for local spreading. We show for interactions between ``Bronze'' and ``Gold" and ``Bronze'' and ``Silver" communities, the detrimental, one-way, effect of non-vaccinated communities to vaccinated ones. When considering the spread of a virus in Barab\'asi-Albert networks, moderately vaccinated ``Silver'' and well-vaccinated ``Gold'' communities are able to resist the spread only for some time as the vaccine efficacy wanes over time. Our work shows that the ``Gold'' communities are the best to keep their infection levels low, however even a small number of unvaccinated ``Bronze'' communities are enough to spiral up infections in moderately and well-vaccinated communities. We find that even a large number of ``Gold'' communities is unable to keep infection levels low or halt the prevalence of the spread. Interestingly, interconnected communities in dense Erd\H{o}s-R\'enyi random networks reach the disease-free state, but not in sparse Watts-Strogatz and Barab\'asi-Albert networks, where the infections spread to all communities in the long term. Based on our results, we conclude that the presence of unvaccinated or partially vaccinated communities in a network, can increase significantly the rate of infected population in other communities. This reveals the importance of maintaining vaccination campaigns globally to defend ourselves against the spread of infectious diseases such as COVID-19.}

\section{Introduction}\label{sec_intro}


The novel strand of Coronavirus (SARS-CoV-2) was identified in Wuhan, Hubei Province in China in December 2019 and it has spread since then everywhere in the world \cite{Miaoetal2017,Who2021b,tang2020estimation,wu2020new,Who2020}. The World Health Organization (WHO) declared it a pandemic on 11 March 2020. It is known that it causes a severe and potentially fatal respiratory syndrome, i.e., COVID-19 which has \cite{wu2020new} impacted heavily on human health and on the socioeconomic status in affected countries. Governments and local authorities imposed counter measures to mitigate its spread, such as wearing face masks, sanitisation, hand-washing, social-distancing, implementation of local and national lockdowns, quarantines, etc. They have all proved quite successful in slowing down the spread of the virus \cite{Peaketal2020}, however, there are unmet challenges \cite{Bar-Zeev2020}, for example when such measures are implemented for long times or repeatedly. Vaccination programs have started being implemented around the world \cite{Who2021a} and the social landscape in affected countries has changed due to the virus and vaccines. Hence it is important to consider broader mitigation measures to halt the spread of COVID-19, not only within communities such as in countries but importantly, among different interconnected communities. This is also because citizens can travel through transportation, flight or other networks and mix with each other. The effect of vaccination in interconnected communities is crucial in understanding the spread of COVID-19 and the development of the pandemic in a global scale.

Mathematical approaches to model and analyse the dynamics of infectious diseases, including COVID-19, have been developed to forecast the trajectory of viruses or depict trends and patterns. For example, the work in \cite{wu2020new,wu2020nowcasting,ranjan2020predictions,Who2020,boccaletti2020modeling,machado2020nonlinear,cooper2020sir,cooper2020dynamic} focuses on the changes of susceptibility, infection rates, deaths and recovered cases from COVID-19, which can help governments and local authorities implement counter measures to reduce infection rates. If measures are not taken to mitigate the spread, the number of infected cases grows exponentially fast with a certain rate of transmission \cite{cooper2020sir,cooper2020dynamic}. Hence the necessity to explain and forecast the trajectory of an infectious virus in interconnected, vaccinated, communities becomes imperative as it can help governments and authorities implement timely, counter measures, prevention and control strategies and to allocate wisely financial and medical resources.

One of the first mathematical models that was used to study the spread of viruses in communities is the susceptible-infected-removed (SIR) model. In particular, Kermack and McKendrick \cite{Kermacketal2020} presented in 1927 one of the first models to forecast an epidemic. According to their model, a population is divided into the susceptible population ($S$) whose individuals can become infected, the infected population ($I$) and the removed population ($R$) who are individuals who have either recovered or have died due to the virus. The SIR model is given by a system of three coupled, nonlinear, ordinary differential equations (ODEs) which describe the dynamics of the disease in time as individuals move from one compartment to another. Since then, the SIR model and modifications \cite{Hethcote1989,Hethcote2000,Hethcote2008,Weiss2013} have been used \cite{LOPEZ2021103746,Ndairou2020} to model the spread of COVID-19 in communities \cite{cooper2020sir,cooper2020dynamic,Anas2020,Kraemer2020,Khoshnawetal2020,Tuiteetal2020,Almeshaletal2020,10.3389/fmed.2020.00171}. The main goal is to predict the duration of the pandemic and to find how interventions such as social distancing, immunisation and vaccination could reduce the number of infected individuals. Importantly, mathematical models can be used to evaluate the effectiveness of control policies against the spreading of infectious diseases, such as COVID-19.

One of the effective ways to immunise communities is through vaccination. Epidemic models that include vaccination are divided into two categories. In the first, the models assume vaccination as a treatment, whereby individuals are moved from the infected to the recovered population \cite{arino2003,Gakkhar2008,peng2019,wang2019}. In the second category, the models consider the vaccinated individuals as members of a separate population \cite{liu2017,peng2016,peng2013,scherer2002}, which is modelled by its own ODE. According to the models in the first category, it is assumed that the population is vaccinated at a constant rate. However, in reality, the vaccination and control strategy is dynamic in time. Hence here we adopt the approach in the second category and model vaccinated populations with a separate ordinary differential equation.

These approaches assume the communities (e.g., countries) are isolated and hence do not interact with their external environment, i.e., with other communities. This means mixing of individuals from different communities cannot occur, not even when vaccination strategies are in place to increase herd immunity in the communities. However, in reality, even when national lockdowns, quarantines or other mitigation measures are in place (e.g., vaccinations), there are still people traveling from one place to another, mixing and coming in contact with people from other communities, potentially spreading the virus. This brings about the necessity to develop mathematical models to study the spread of a virus in interconnected communities in which the percentage of the vaccinated population and vaccine efficacy vary.

Here, we propose a mathematical model to study the spread of a virus in interconnected communities and the effects of different levels of vaccination, employing ideas from complex systems and complex networks \cite{Sayama2015}. As in reality, infected individuals or individuals that become infected during their trip, can travel from one place to another and spread a virus, here we focus on this case and allow only infected individuals to travel through a network of interconnected communities. We explain in the supplementary material why we have made this choice. We model the interconnected communities as a network where the communities (e.g., countries) are the nodes and their interactions (i.e., travelling among communities) are the connections, which are assumedly undirected for simplicity. We consider different types of connectivity networks, such as Erd\H{o}s-R\'enyi (ER) random \cite{Erdosetal1959}, Watts-Strogatz (WS) small-world \cite{Wattsetal1998} and Barab\'asi-Albert (BA) scale-free \cite{RevModPhys.74.47} networks. On each node of the network we consider an epidemic model with vaccination strategy (SVIR) based on the classic SIR model \cite{Kermacketal2020}, where nodes are connected to other nodes via the Laplacian of the adjacency matrix of the network to model local, diffusive, spreading. In this context, $V$ stands for the vaccinated population in the community and $S$, $I$ and $R$ are the susceptible, infected and recovered populations in the community, respectively. This results in a system of ODEs coupled via the Laplacian matrix of the network. As we consider undirected connections only, the adjacency matrices are symmetric and binary, whereby 0 means no connection and 1 a connection.

Our work is motivated by the work in \cite{Tornatoreetal2014,LIU20081}, where SVIR epidemic models with vaccination strategies and a SVIR epidemic model with stochastic perturbations were studied, and provides a mathematical approach to model the effect of different levels of vaccination in susceptible, infected and removed populations in interconnected communities. A basic version of our numerical code used in the paper is shared on GitHub \cite{githubSVIRNet2020} to help anyone interested to study the spread of viruses in interconnected communities. The paper is also accompanied by a supplementary material, where we present additional results on the spread of a virus on ER random, WS small-world and BA scale-free networks and on network properties. In particular, we present results for communities sorted based on the node-degrees in a BA scale-free network, results for the case where only susceptible populations can travel through a network and at the end, we discuss some of the spectral properties of the networks used herein.

The paper is organised as follows: Section \ref{sec_methodology} discusses our approach in modelling mathematically the spread of a virus in interconnected communities. In Sec. \ref{sec_parameters_scenarios} we explore a number of vaccination scenarios in interconnected communities, assuming that initially, all populations are susceptible. We discuss our main results in Sec. \ref{sec_main_results} and in Sec. \ref{sec_discussion_conclusions}, we conclude our work, discussing the outcomes of our analysis in a broader context, in view of the evidence that has been collected on the spread of COVID-19 worldwide.

\section{Modelling the spread of a virus in interconnected communities}\label{sec_methodology}

\subsection{From SIR to SVIR modelling of a virus}\label{sec_SIR_SVIR_models}


The modelling approach we are taking here to study the spread of a virus (e.g., COVID-19) in interconnected communities is based upon the classic susceptible-infected-removed (SIR) model \cite{weiss2013SIR}, which is a simple compartmental model to study the spread of viruses in communities \cite{weiss2013SIR,Kermacketal2020,amaro2021global}. The interconnected communities are modelled by a network of $N$ nodes, whereby each node is a community, resulting in $N$ communities connected through the Laplacian of the adjacency matrix of the network, for local, diffusive, spreading. In the SIR model, each community is split into three compartments, i.e., into the susceptible $S$, infected $I$ and removed $R$ populations that evolve in time $t$. In this context, the total population $M=S+I+R$ is constant. In particular, $M$ is split into the:
\begin{enumerate}
\item {Susceptible population, $S$: These are the individuals who are not infected, but can become infected. A susceptible individual may become infected or remain susceptible. As the virus spreads from its source or new sources spring up in the community, more individuals become infected, thus the susceptible population decreases in time.}
\item{Infected population, $I$: These are the individuals who have already been infected by the virus and can transmit it to the susceptible individuals. An infected individual may remain infected, and can be removed from the infected population to recover or die.}
\item{Removed population, $R$: These are the individuals who have either recovered from the virus and are assumed to be immune or have died.}
\end{enumerate}

The dynamics of the classic SIR model \cite{weiss2013SIR} is determined by the system of ODEs
\begin{align}\label{eq_classic_SIR_Model}
\frac{dS}{dt}&=-\beta SI,\nonumber\\
\frac{dI}{dt}&=\beta SI-\lambda I,\\
\frac{dR}{dt}&=\lambda I,\nonumber
\end{align}
where $\beta$ is the probability of infections per day, meaning that each susceptible individual infects randomly $\beta$ individuals every day and $\lambda$, the fraction of infected individuals that are transferred to the removed population. In this context, $\beta$ and $\lambda$ are constants. The SIR model \eqref{eq_classic_SIR_Model} is derived assuming that: (a) the members of the susceptible and infected populations are homogeneously distributed in space and time, (b) an individual removed from the infected population has lifetime immunity, (c) the total population $M$ is constant in time, (d) the time-scale of the SIR model is short enough so that births and deaths, other than deaths caused by the virus, can be neglected and (e) the number of deaths from the virus is small compared with the living population.

Even though the classic SIR model \eqref{eq_classic_SIR_Model} is a simple, compartmental, model that can describe the spread of a virus in a community, it does not account for a vaccinated population and thus, cannot describe its influence on the dynamics of $S$, $I$ and $R$ in the community. Here, we are interested in studying the spread of a virus, e.g., COVID-19, and the effect of different levels of vaccination, in interconnected communities. The authors in \cite{Tornatoreetal2014} present the susceptible-vaccinated-infected-removed (SVIR) model
\begin{align}\label{eq_original_single_SVIR_Model}
\frac{dS}{dt}&=\mu-\beta SI-(\mu +\phi) S,\nonumber\\
\frac{dV}{dt}&=\phi S-\rho\beta IV-\mu V,\\
\frac{dI}{dt}&=\beta SI+\rho\beta IV-(\lambda +\mu) I,\nonumber\\
\frac{dR}{dt}&=\lambda I-\mu R,\nonumber
\end{align}
which is based on the classic SIR model \eqref{eq_classic_SIR_Model}. Model \eqref{eq_original_single_SVIR_Model} describes the spread of a disease when a vaccination program is in effect, assuming the vaccine does not lose its efficacy. Further on, it supposes that in the unit of time, a fraction $\phi$ of the susceptible population, $S$, is vaccinated. The vaccination may reduce but not completely eliminate susceptibility to infection, so the model includes a factor $\rho$, where $0\leq\rho\leq1$, in the contact rate of vaccinated members, $V$, with $\rho= 0$ corresponding to the case where the vaccine has perfect efficacy and $\rho = 1$ to the case where the vaccine has no effect at all. The model also supposes that the immunity is permanent so that a fraction $\lambda$ of infected individuals, $I$, returns to the removed population, $R$, that births occur with the same constant rate $\mu$ of deaths and that all newborns enter in the susceptible population, $S$. Hence $\mu$, $\lambda$, $\phi$ and $\beta$ are all positive, real numbers.

Motivated by the work in \cite{Tornatoreetal2014,LIU20081}, where SVIR epidemic models with vaccination strategies and SVIR epidemic model with stochastic perturbations are studied, and driven by the timely question of what is the effect of different levels of vaccination in interconnected communities and in their susceptible, infected and removed populations, we introduce the following system of coupled ODEs
\begin{align}\label{svir_uncoupled}
\frac{dS}{dt}&=-\beta SI-\phi S+\delta R,\nonumber\\
\frac{dV}{dt}&=\phi S-\rho\beta IV,\nonumber\\
\frac{dI}{dt}&=\beta SI+\rho\beta IV-\lambda I,\\
\frac{dR}{dt}&=\lambda I-\delta R,\nonumber
\end{align}
for the spread of a virus in a single community. System \eqref{svir_uncoupled} is derived from system \eqref{eq_original_single_SVIR_Model} for $\mu=0$, where we have also added and subtracted the term $(\delta R)$ in the first and from the last equation, respectively. Setting $\mu=0$ assumes there are no births and no deaths in the community, other than deaths due to the virus, as in the classic SIR model \eqref{eq_classic_SIR_Model}. Parameter $\delta$ is the constant rate for the loss of immunity and hence $1/\delta$, the mean immune period \cite{VARGASDELEON20111106}. In the special case where $\delta = 0$, the immunity is permanent and there is no return from the recovered population, $R$, to the susceptible population, $S$, in which case the resulting model is a SIR model. Clearly, in model \eqref{svir_uncoupled}, $S+V+I+R=M$, so that the total population remains constant in time. System \eqref{svir_uncoupled} can also be derived from the classic SIR model \eqref{eq_classic_SIR_Model}, by adding the second equation of system \eqref{svir_uncoupled} for the vaccinated population and the terms $(-\phi S+\delta R)$, $(-\delta R)$ in the first and third equations in the classic SIR model \eqref{eq_classic_SIR_Model}. Parameter $\lambda$ is the rate of recovered individuals in the community, $\beta$ the transmission rate of susceptible to infected individuals, $\phi$ the percentage of the vaccinated population in the susceptible population $S$ and $0\leq \rho\leq 1$, the efficacy of vaccination in the community. In this context, $\rho=0$ means perfect vaccine efficacy, whereas $\rho=1$ means the vaccine has no effect at all. Thus, taking $\phi=\rho=0$ means the community is not vaccinated at all.

\begin{figure}[!ht]
\centering
\includegraphics[height=5.7cm,width=0.48\textwidth]{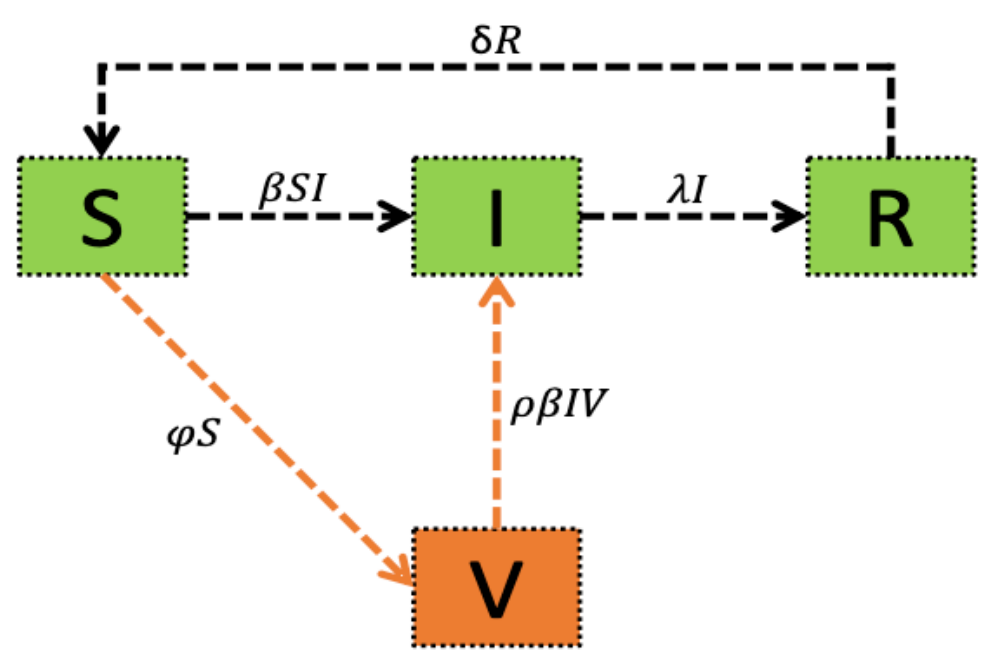}
\caption{The transfer or population-flux diagram of SVIR model \eqref{svir_uncoupled}. The diagram shows how the susceptible $S$, vaccinated $V$, infected $I$ and removed $R$ populations in a single community are influencing each other. A subpopulation of vaccinated individuals, $V$, is added to the standard SIR model \eqref{eq_classic_SIR_Model}. Parameter $\phi$ is the rate at which susceptible individuals enter the vaccination process. The product $(\rho\beta)$ is the transmission rate for vaccinated individuals to be infected before gaining immunity, $\delta$ the constant rate for the loss of immunity, $\lambda$ the rate of recovered individuals in the community and $\beta$, the transmission rate of susceptible to infected individuals. The arrows show the direction of the flow of populations in the community and their labels, the rates or percentages of the flow.}\label{fig_SVIR_uncoupled_flowchart}
\end{figure}

The transfer or population-flux diagram in Fig. \ref{fig_SVIR_uncoupled_flowchart} shows how the four populations in system \eqref{svir_uncoupled} are influencing each other within a community. As $\phi$ is the percentage of susceptible individuals being vaccinated, they will obtain vaccine-induced immunity during or after the vaccination process. Model \eqref{svir_uncoupled} does not distinguish between natural and vaccine-induced immunity as the latter can also last for some time. Furthermore, the model assumes that before obtaining immunity, the vaccinated individuals still have the possibility of infecting other individuals with a disease transmission rate $(\rho\beta)$, while contacting with infected individuals. In this context, $\rho\beta<\beta$ as the vaccinated individuals may, for example, have developed partial immunity or they may be able to minimise their exposure to infected individuals.

The equations of the classic SIR model \eqref{eq_classic_SIR_Model} and systems \eqref{eq_original_single_SVIR_Model} and \eqref{svir_uncoupled} define three systems of coupled nonlinear ODEs, and thus finding analytical solutions in closed form, given by known mathematical functions, is difficult \cite{Braueretal2019}. Even though one can derive the analytical solution in an implicit form for system \eqref{eq_classic_SIR_Model}, the process to find it, is complicated and there are limitations in practical applications \cite{Harkoetal2014}. Thus, a common approach is to solve these systems numerically. In our simulations throughout the paper, we have used the numerical integrator {\it ode45} in Matlab, which is a single-step solver, based on an explicit Runge-Kutta (4,5) formula, the so-called Dormand - Prince pair \cite{Dormandetal1980}. It uses a variable integration-time step, comparing methods of orders four and five to estimate the error. In the following, we will assume that all quantities are dimensionless and that we are interested in performing a qualitative study.

Next we focus on the dynamics of the modified SVIR model \eqref{svir_uncoupled} and study the behaviour of its solutions in the neighbourhood of its equilibrium points, performing an equilibrium stability analysis.

\subsection{Equilibrium stability analysis of the SVIR and SIR models with waning immunity}\label{subsec_esa_single_SVIR_model}

Here we shed light on the dynamics of systems \eqref{svir_uncoupled} performing an equilibrium stability analysis to study the stability of its trajectories in the neighbourhood of its equilibrium points. The system has the disease-free equilibrium point (DFE) $E_0=(S_0,V_0,I_0,R_0)=(0,V_0,0,0)$ for any $V_0$. As in this model, the sum of the time-derivatives of $S$, $V$, $I$ and $R$, $\dot{S}+\dot{V}+\dot{I}+\dot{R}=0$, we have that $S+V+I+R=M$, where $M$ is the constant total population. Hence the total population remains constant in time. We suppose, without loss of generality, that $M=1$, which results in $E_0=(0,1,0,0)$, where $V_0=M=1$. The eigenvalues of the Jacobian of system \eqref{svir_uncoupled} at $E_0$ are $0$, $-\phi$, $-\delta$ and $\rho\beta-\lambda$. Therefore, $E_0$ is stable if $\rho\beta-\lambda<0$ as $\phi$ and $\delta$ are positive. Setting $\mathcal{R}_0=\frac{\rho\beta}{\lambda}$, if $\mathcal{R}_0<1$, $E_0$ is stable, whereas if $\mathcal{R}_0>1$, $E_0$ is unstable as there is one positive eigenvalue, i.e., $\rho\beta-\lambda>0$. Hence, system \eqref{svir_uncoupled} has a stable DFE for $\mathcal{R}_0<1$ and an unstable one for $\mathcal{R}_0>1$. In this framework, $\mathcal{R}_0$ is the basic reproduction number of system \eqref{svir_uncoupled} and determines the stability of DFE.

Furthermore, when $E_0$ is unstable, another equilibrium point is obtained where
\begin{align*}
S^*&=\frac{\lambda I^*}{\beta I^*+\phi},\\
V^*&=\frac{\lambda\phi}{\rho \beta (\beta I^*+\phi)},\\
R^*&=\frac{\lambda I^*}{\delta},
\end{align*}
with $I^*$ being the positive root of the quadratic
\begin{equation}
\mathcal{A}I^2+\mathcal{B}I+\mathcal{C}=0,\label{quadratic_I_eq}
\end{equation}
resulting from $S^*+V^*+I^*+R^*=M$, where
\begin{align*}
\mathcal{A}&=\rho \beta^2(\delta+\lambda),\\
\mathcal{B}&=\rho \beta \phi (\delta+\lambda)+\rho\beta\delta(\lambda-\beta),\\
\mathcal{C}&=\phi\delta(\lambda-\rho\beta)=\phi\delta\lambda(1-\mathcal{R}_0).
\end{align*}
In particular, since $\mathcal{A}>0$ as $\beta$, $\rho$, $\phi$, $\lambda$ and $\delta$ are all positive, parabola \eqref{quadratic_I_eq} opens upward, having 0, 1 or 2 roots. In the case of 2 roots, the quadratic equation \eqref{quadratic_I_eq} has one negative and one positive root. We keep the positive root $I=I^*$ to use in $S^*$, $V^*$ and $R^*$ as $I$ is a population, so $I(t)\geq 0$ for all times $t$.

Next we investigate the dynamics of $S$, $I$ and $R$ populations in the absence of vaccination where the mean immune period is $1/\delta$ \cite{VARGASDELEON20111106}, hence we remove the second ODE for vaccination from system \eqref{svir_uncoupled} and set $\phi=\rho=0$ to end up to the reduced system
\begin{align}\label{eq_Bronze_type_Model}
\frac{dS}{dt}&=-\beta SI+\delta R,\nonumber\\
\frac{dI}{dt}&=\beta SI-\lambda I,\\
\frac{dR}{dt}&=\lambda I-\delta R,\nonumber
\end{align}
which is essentially the classic SIR model \eqref{eq_classic_SIR_Model} with waning immunity, assuming $\delta>0$. System \eqref{eq_Bronze_type_Model} has the DFE $P_0=(S_0,I_0,R_0)=(1,0,0)$ and its basic reproduction rate is $\mathcal{R}_{00}=\frac{\beta}{\lambda}$. The eigenvalues of its Jacobian at $P_0$ are $0$, $-\delta$, and $\beta-\lambda$. Therefore, $P_0$ is stable if $\beta-\lambda<0$ or $\mathcal{R}_{00}<1$ and unstable if $\mathcal{R}_{00}>1$. In the latter case, there is another equilibrium point $P^*=(S^*,I^*,R^*)=\Big	(\frac{\lambda}{\delta},\frac{\delta(\mathcal{R}_{00}-1)}{\mathcal{R}_{00}(\delta+\lambda)},\frac{\lambda(\mathcal{R}_{00}-1)}{\mathcal{R}_{00}(\delta+\lambda)}\Big)$.

\begin{figure*}[!ht]
\centering
\includegraphics[height=5.7cm,width=0.99\textwidth]{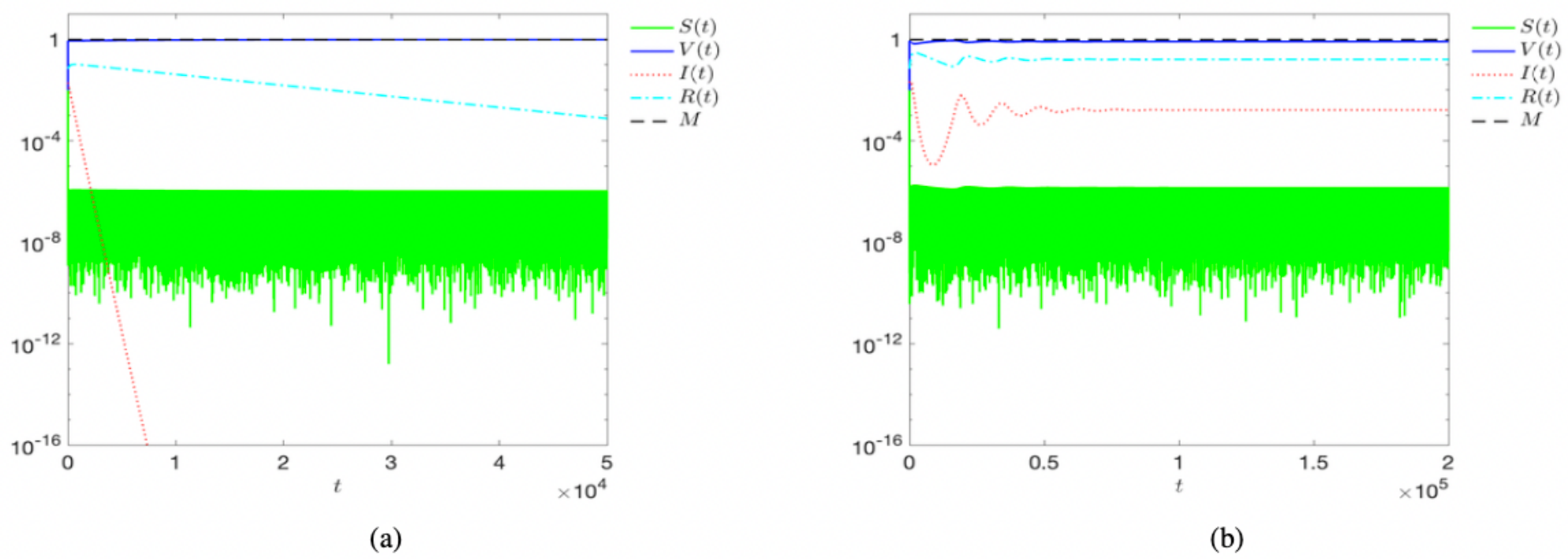}
\caption{Numerical solution to system \eqref{svir_uncoupled} for two sets of parameters (see text) which result in $\mathcal{R}_0<1$ and $\mathcal{R}_0>1$ in panels (a) and (b), respectively. In panel (a), the system converges to the stable DFE $E_0=(S^*,V^*,I^*,R^*)=(0,1,0,0)$ and in panel (b), to the stable equilibrium point given in Eq. \eqref{eq_fp_R0=1.2}. Note that in panel (a), $\mathcal{R}_0=0.6$ and that in panel (b), $\mathcal{R}_0=1.2$. The black dashed line corresponds to $S+V+I+R=M=1$.}\label{fig_R0}
\end{figure*}

Next, we simulate numerically system \eqref{svir_uncoupled} in Fig. \ref{fig_R0} for two sets of parameters that result in $\mathcal{R}_0<1$ (panel (a)) and $\mathcal{R}_0>1$ (panel (b)). First, we choose $(\beta,\rho,\phi,\lambda,\delta)=(0.02,0.3,80,0.01,0.0001)$ (what we call a ``Gold'' community in Subec. \ref{subsec_study_cases}), that results in $\mathcal{R}_0=0.6<1$ and to trajectories approaching the stable DFE at $(0,1,0,0)$ as can be seen in panel (a). In the second case shown in panel (b), we consider the parameter values $(\beta,\rho,\phi,\lambda,\delta)=(0.02,0.6,40,0.01,0.0001)$ (what we call a ``Silver'' community in Subec. \ref{subsec_study_cases}) that result in $\mathcal{R}_0=1.2>1$, in which case $E_0$ is unstable and the trajectories approach the equilibrium point
\begin{equation}
(S^*,V^*,I^*,R^*)\approx(4.125\;10^{-7}, 0.833, 1.65\;10^{-3}, 0.165).\label{eq_fp_R0=1.2}
\end{equation}
This equilibrium point is stable as the eigenvalues are approximately equal to $-40,-5.99\;10^{-5}+4.43\;10^{-4}i, -5.99\;10^{-5}-4.43\;10^{-4}i, 0$, where $i$ is the imaginary unit in the complex plane (i.e., $i^2=-1$). We note that the equilibrium point in Eq. \eqref{eq_fp_R0=1.2} is not a DFE, as $I$ converges to $I^*\approx1.65\;10^{-3}$, in contrast to $E_0$, where $I_0=0$.

Returning to system \eqref{eq_Bronze_type_Model}, considering for example the set of parameters $(\beta,\lambda,\delta)=(0.02,0.01,0.0001)$ results in $\mathcal{R}_{00}=2$. In this case, the system has the unstable DFE $P_0=(S_0,I_0,R_0)=(1,0,0)$ with the eigenvalues being approximately equal to $0,-0.0001,0.01$ and the stable equilibrium $P^*=(S^*,I^*,R^*)\approx(0.5,4.95\;10^{-3}, 0.495)$ with the eigenvalues being approximately equal to $-0.005+0.0087i, -0.005-0.0087i,0$, where $i$ is the imaginary unit in the complex plane (i.e., $i^2=-1$).

\subsection{A network approach for the spread of a virus in interconnected communities}\label{subsec_SVIR_coupled_model}

Building on the SVIR model \eqref{svir_uncoupled}, we consider here $N$ interconnected communities, where each community $i=1,\dots,N$ is divided into four compartments $S$, $V$, $I$ and $R$, i.e., into the susceptible, vaccinated, infected and recovered populations, respectively. Since in reality, infected individuals or individuals that become infected during their trip, can travel from one place to another and spread a virus, we will assume in our model that only infected individuals can travel through a network of interconnected communities. We explain more about this in Sec. III in the supplementary material. The dynamics of the communities then is described by the system of coupled ODEs
\begin{align}\label{SVIR_coupled_system_ODEs}
\frac{dS_i}{dt}&=-\beta_iS_iI_i-\phi_iS_i+\delta_iR_i,\nonumber\\
\frac{dV_i}{dt}&=\phi_iS-\rho_i\beta_iI_iV_i,\nonumber\\
\frac{dI_i}{dt}&=\beta_iS_iI_i+\rho_i\beta_iI_iV_i-\lambda_iI+\alpha\sum_{j=1}^NL^I_{ij}I_j,\\
\frac{dR_i}{dt}&=\lambda_i I_i-\delta_iR_i,\nonumber
\end{align}
where $L^I$ is the Laplacian of the adjacency (or connectivity) matrix $A^I$ that describes the connectivity of the communities. Hence
\begin{equation}
L^I=K^I-A^I,\label{laplacian_matrix}
\end{equation}
where $K^I$ is the node-degree matrix of $A^I$. Equation \eqref{laplacian_matrix} then implies that
\begin{equation}\label{eq_laplacian_sum_over columns}
\sum_{j=1}^NL^I_{ij}=0,
\end{equation}
where $i=1,\ldots,N$. Since $I_i\in[0,1]$, the coupling term in the third equation in system \eqref{SVIR_coupled_system_ODEs} gives rise to a constant population $M=\sum_{i=1}^N (S_i+V_i+I_i+R_i)=N$. In other words, the total population in system \eqref{SVIR_coupled_system_ODEs} is constant in time, and in particular, it is equal to $N$, that is to the number of communities in the network. The coupling term denotes the diffusive coupling among the infected populations, whereby the connectivity strength is denoted by $\alpha\geq0$. 

The communities at the nodes of the network may be vaccinated, partially vaccinated or unvaccinated. First, we divide each community $i=1,\dots,N$ into the susceptible $S_i$, vaccinated $V_i$, infected $I_i$ and recovered $R_i$ populations, which are functions of time, $t$. The way populations $S_i$, $V_i$, $I_i$ and $R_i$ in community $i$ are influencing each other is shown schematically in Fig. \ref{fig_SVIR_uncoupled_flowchart}.

In our numerical approach, we assume that initially (i.e., at $t=0$) all individuals in a community, either that is a ``Bronze'', ``Silver'' or ``Gold' community (see Table \ref{table_1}), are susceptible to the virus and that there are no vaccinated, infected and recovered individuals. This leads to the initial conditions $S_i(0)=1$, $V_i(0)=0$, $I_i(0)=0$ and $R_i(0)=0$ for all $i$, used throughout the paper.


\section{Modelling vaccination scenarios}\label{sec_parameters_scenarios}

In modelling the dynamics of infections in vaccinated, interconnected communities using system \eqref{SVIR_coupled_system_ODEs}, the coupling term $\alpha$ models the easiness of travel through the network of the $N$ interconnected communities. For example, $\alpha=0$ corresponds to $N$ fully locked down (or disconnected) communities, whereby travelling from one to another is not possible. Hence increasing positive values of $\alpha$ correspond to travelling easier through the network.

Given that governments and local authorities impose partial or full lockdowns or other measures to hinder or ban people from travelling and mixing with other groups of people as a means to control the spread, we will assume in our study that $\alpha$ is small. Our results show that in sparse (i.e., very low network density) WS small-world and BA scale-free networks, large values of $\alpha$ lead eventually to a homogenisation of the spread of infections in the communities (see Fig. \ref{fig_a2=0.01_thermalisation}), assuming initially the populations are susceptible. Our results also show that for dense-enough ER random, WS small-world and BA scale-free networks, the dynamics reach the disease-free state (see Fig. \ref{fig_sec_network_properties_and_the_spread_of_a_virus}). To model and study the effect of incomplete lockdowns in interconnected communities, we will start by setting $\alpha=10^{-8}$, to account for a small coupling term, enough to give rise to interesting phenomena highlighting the dramatic influence of non-vaccinated communities to other communities.

\subsection{Parameters}\label{subsec_study_cases}


In the following, we will study the spread of an infectious disease, e.g., COVID-19, for different vaccination scenarios by splitting the $N$ communities into three, not necessarily equally-sized, groups of sub-communities, to model different vaccination levels. We consider ``Gold'' communities with a high percentage of vaccinated individuals and high vaccination efficacy, ``Silver'' with a moderate percentage of vaccinated individuals and vaccination efficacy and ``Bronze'' with no vaccinated individuals at all. In the case of unvaccinated ``Bronze'' communities, the efficacy of the vaccination does not play actually any role as no one in the community is vaccinated, however for numerical reasons it is set equal to 0, meaning that $\rho=0$ for the ``Bronze'' communities. Our choice of types of communities covers vaccination scenarios ranging from excellent (``Gold'') to moderate (``Silver'') to poor (``Bronze'') vaccination policies. We summarise the parameter values used throughout the paper for the three types of communities in Table \ref{table_1}.

\begin{table}[h!]
\begin{center}
\begin{tabular}{|c|c|c|c|c|c|}
\hline
& \multicolumn{5}{|c|}
{Parameter values}\\ \cline{2-6} 
\hline 
Type of community &$\beta$&$\phi$ (\%)&$\rho$&$\lambda$&$\delta$\\ \hline 
$\text{``Bronze''}$&$0.02$&$0$&$0^*$&$0.01$&$0.0001$\\
\hline
$\text{``Silver''}$&$0.02$&$40$&$0.6$&$0.01$&$0.0001$\\
\hline
$\text{``Gold''}$&$0.02$&$80$&$0.3$&$0.01$&$0.0001$\\
\hline
\end{tabular}
\caption{The parameters used in system \eqref{SVIR_coupled_system_ODEs} to model ``Bronze'', ``Silver'' and ``Gold'' communities throughout the paper. We note that $\phi$ is expressed as the percentage of the susceptible population that is vaccinated and that $0^*$ means that in the case of the unvaccinated ``Bronze'' community, the efficacy of the vaccination does not play any role as no one in the community is vaccinated, however for numerical reasons it is set equal to 0.}\label{table_1}
\end{center}
\end{table}


Based on the values in Table \ref{table_1}, the ``Gold'' communities are considered the most well-vaccinated, the ``Silver'', the second best well-vaccinated and the ``Bronze'' the worst vaccinated. In the latter case, we assume that vaccination is not available, hence the choice $\phi =0$. Parameter $\rho$ is the efficacy of vaccination in the communities and hence $\rho=0$ means the vaccine is fully effective and $\rho=1$ that the vaccine has no effect at all. Thus, taking $\phi=\rho=0$ means the community is not vaccinated. Here we consider that the ``Gold'' communities are the best vaccinated with very high efficacy, i.e., $\rho=0.3$ and the ``Silver'' communities the second-best well-vaccinated communities using $\rho=0.6$. For the ``Bronze'' communities, we assume that they cannot roll out vaccination policies, and hence set $\phi=\rho=0$ for these communities. Finally, we consider that the rest of the parameters, i.e., $\beta$, $\lambda$ and $\delta$ are the same for all types of communities as reported in Table \ref{table_1}.

\subsection{Networks}\label{subsec_networks}

\begin{figure}[!ht]
\centering
\includegraphics[height=17.1cm,width=0.48\textwidth,angle =0]{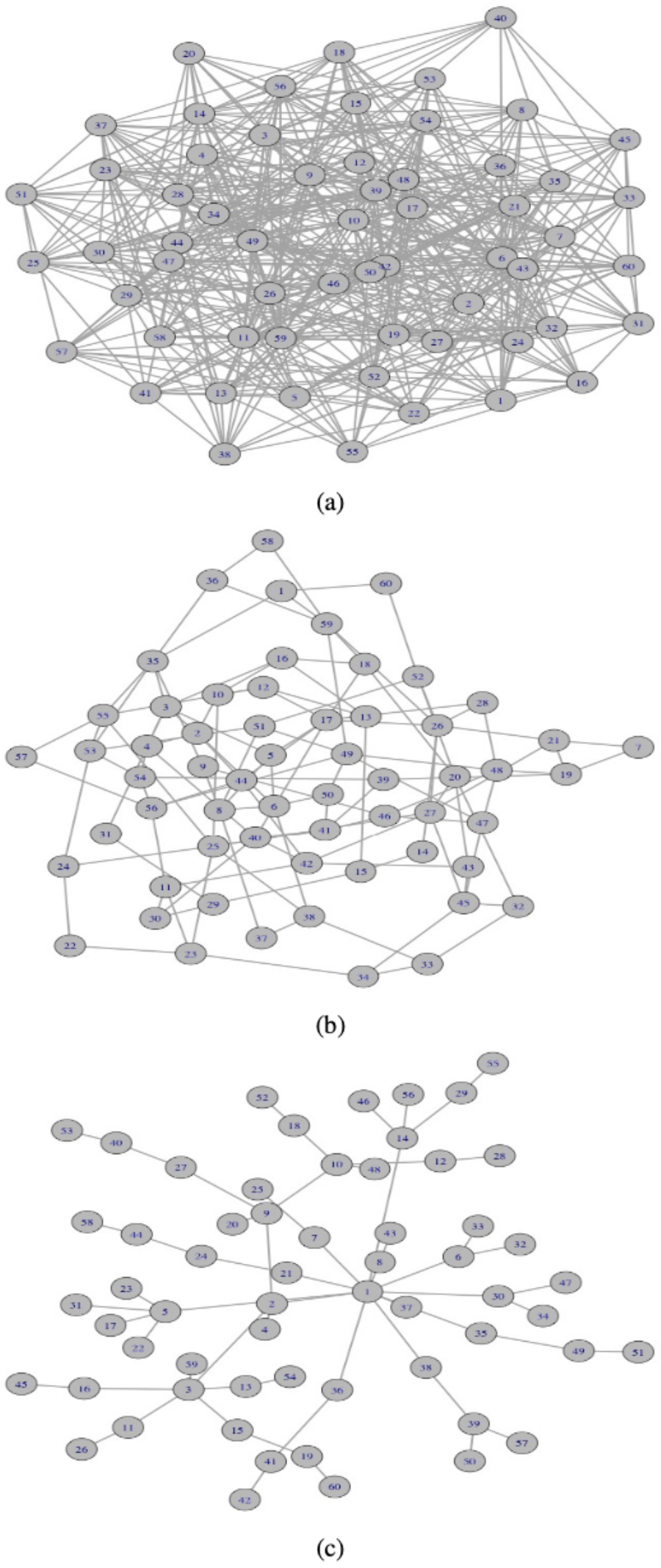}
\caption{Examples of an ER random network with rewiring probability 0.3 (resulting in network density 0.3) in panel (a), a WS small-world network in panel with network density 0.07 in panel (b) and a BA scale-free network with network density 0.033 in panel (c). In all cases, $N=60$ nodes. We note that the BA network in panel (c) contains hub nodes which is not the case with the networks in panels (a) and (b).}\label{fig_ERN_WSN_BN_60_nodes}
\end{figure}

In the following, we consider networks composed of vertices or nodes and edges or links \cite{Newman2010}. The vertices or nodes play the role of communities and the edges, the role of links (or interconnections) connecting the communities. For simplicity, we assume that the links are undirected, meaning that if community $i$ is connected to community $j$, then community $j$ is connected to community $i$ at the same time. We consider three exemplar network topologies, i.e., ER random \cite{Erdosetal1959}, WS small-world \cite{Wattsetal1998} and BA scale-free \cite{RevModPhys.74.47} networks of $N=60$ interconnected communities (i.e., nodes). To compute the networks and corresponding adjacency (or connectivity) matrices $A^I$ used in our study, we have employed the igraph package in R. From these adjacency matrices $A^I$, the corresponding Laplacian matrices $L^I$ were computed using Eq. \eqref{laplacian_matrix}, which were then fed into system \eqref{SVIR_coupled_system_ODEs}. In particular, in Fig. \ref{fig_ERN_WSN_BN_60_nodes} we consider an example of an ER random network with rewiring probability $p=0.3$ (resulting in network density 0.3) in panel (a), a WS small-world network with network density 0.07 in panel (b) and a BA scale-free network with network density 0.033 in panel (c). In Fig. \ref{fig_sec_network_properties_and_the_spread_of_a_virus}, we also consider WS small-world and BA scale-free networks with density 0.3, the same with the ER random network. The network density is given by the ratio of actual connections to potential connections in the network and ranges between 0 and 1. Hence, the ER random network in panel (a) is denser than the other two networks.

In ER random networks, all graphs on a fixed vertex set with a fixed number of edges are equally likely. A random network consists of $N$ nodes where each node pair is connected with a predefined rewiring probability $p$. It is a very simple model where every possible edge is created with the same rewiring probability $p$. A small-world network is a network where the typical distance $L$ between two randomly chosen nodes (the number of steps required) grows proportionally to the logarithm of the number of nodes $N$ in the network. For example, in a small-world network of people, any two people in the network can reach each other through a short sequence of acquaintances. A common feature of real world networks is the presence of hubs, that is of a few nodes that are highly connected to other nodes in the network. Hubs result in long tails in the network degree distribution, indicating the presence of nodes with a much higher degree than most other nodes. The fact that in many real world networks a small number of highly connected hubs exist has important consequences for how, for example, information and diseases travel through the network. This is relevant to epidemiology, public relations, marketing, etc. Scale-free networks are a type of network characterised by the presence of large hubs. A scale-free network is one with a power-law degree distribution. Several natural and human-made systems, including the Internet, the world wide web, citation networks, and some social networks are thought to be approximately scale-free and certainly contain few nodes with high degree as compared to the other nodes in the network. The BA model \cite{RevModPhys.74.47} is an algorithm for generating random scale-free networks using a preferential attachment mechanism. Many observed networks (at least approximately) fall into the class of scale-free networks, meaning that they have power-law (or scale-free) degree distributions, while ER random and WS small-world networks do not exhibit power laws. The BA model incorporates two important general concepts: growth and preferential attachment. Both growth and preferential attachment exist widely in real networks \cite{Newman2010}.

We start in the next section studying the dynamics of an unconnected ``Bronze''', ``Silver'' and ``Gold'' community, meaning the case where $N=3$ and $\alpha=0$ in system \eqref{SVIR_coupled_system_ODEs} with the rest of the parameters given in Table \ref{table_1}.

\section{Main results}\label{sec_main_results}

\subsection{Three unconnected ``Bronze'', ``Silver'' and ``Gold'' communities}\label{subsec_3_unconnected_B_S_G_communities}

Before considering the case of a big number of interconnected communities and the spread of the virus in large-size networks, we will examine the dynamics of the spread in infections $I(t)$ in system \eqref{SVIR_coupled_system_ODEs} for a trio of unconnected ``Bronze''', ``Silver'' and ``Gold'' communities. This means we start by studying the case where $N=3$ and $\alpha=0$ in system \eqref{SVIR_coupled_system_ODEs} with the rest of the parameters given in Table \ref{table_1}.

\begin{figure}[!ht]
\centering
\includegraphics[height=5.7cm,width=0.48\textwidth]{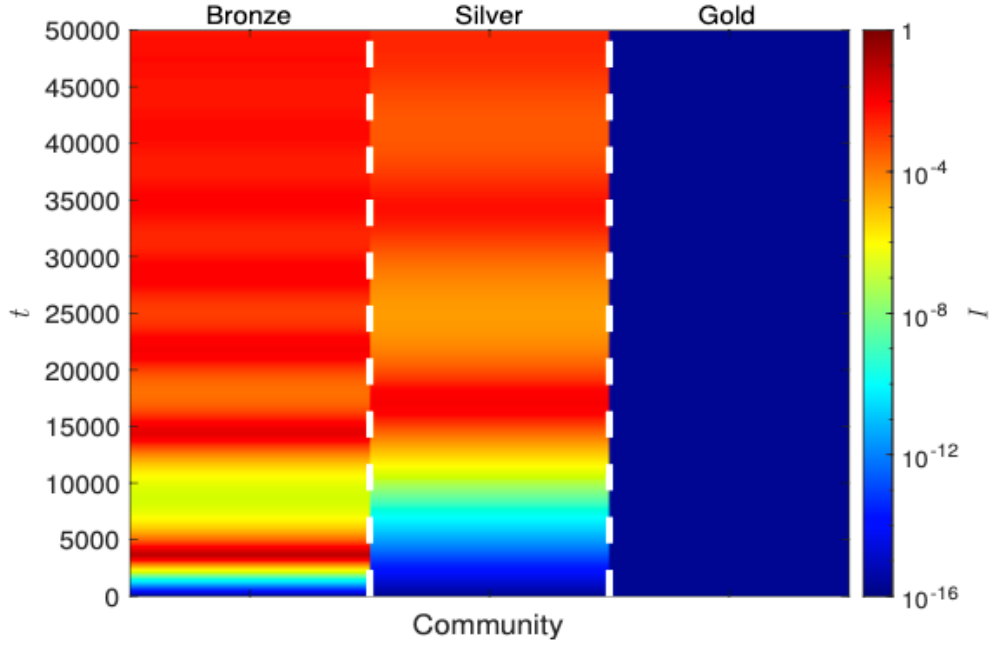}
\caption{Spatiotemporal plot of the evolution of infections, $I$, for three unconnected ``Bronze'', ``Silver'' and ``Gold'' communities modelled by system \eqref{SVIR_coupled_system_ODEs}, where $\alpha=0$. The horizontal axis denotes the community index and the vertical the time $t$. The colour bar encodes the infections $I_i,\;i=1,2,3$ in a logarithmic scale to depict small variations in infections near 0 as $I$ takes values in the interval $[0,1]$. The vertical white, dashed lines delineate the three communities.}\label{fig_three_non_connected_BSG}
\end{figure}

Figure \ref{fig_three_non_connected_BSG} shows the time-evolution of the, non-interacting (i.e., $\alpha=0$), infected populations, $I_i$, $i=1,2,3$, of system \eqref{SVIR_coupled_system_ODEs} where $i=1$ corresponds to the ``Bronze'' community, $i=2$ to the ``Silver'' community and $i=3$ to the ``Gold'' community. In particular, Fig. \ref{fig_three_non_connected_BSG} is a spatiotemporal plot where the horizontal axis denotes the community index and the vertical the time $t$. The colour bar encodes the infections $I_i$ in a logarithmic scale to depict small variations in infections near 0 as $I$ takes values in the interval $[0,1]$. As we can see, infections in the ``Bronze'' community quickly soar and reach close to the maximum at 1 as its population is unvaccinated and is not in contact with another vaccinated population to influence it. It starts in blue meaning it is a fully healthy community with almost no infections. However, since the community is not vaccinated, a rapid and intense first wave of infections at $t\approx 4000$ occurs, followed by a second and third waves at $t\approx 15000$ and $t\approx 22000$, respectively. As it is not vaccinated, a cascade of secondary infection waves follow up leading to an alarming number of infections after about $t\approx 15000$, depicted by red in the spatiotemporal plot. As the scale in the colour bar is logarithmic, deep red at $t\approx 4000$ signifies a tenfold increase in the order of magnitude of infections in the ``Bronze'' community.

In contrast, the situation in infections in the ``Silver'' community is better as $\phi_2=40\%$ of the population is vaccinated and the vaccination efficacy is $\rho_2=0.6$ with 0 corresponding to maximum efficacy (see Table \ref{table_1}). Expectedly, the best performing community in the sense of the smallest number of infections is the ``Gold'' community with the highest percentage of vaccinated population ($\phi_3=80\%$) and vaccination efficacy ($\rho_3=0.3$). In this case, $I_3$ attains very small values of infections around $10^{-16}$. It is evident in the spatiotemporal plot of the ``Bronze'' community that the infections spiral up very quickly in time as the community is not vaccinated, thus the virus prevails very quickly. In the case of the ``Silver'' community, the effect of vaccination does not hold for ever (blue region) as the rate for the loss of immunity from the vaccination $\delta$ is set to 0.0001 and the percentage of vaccinated individuals is $\phi_2=40\%$. Hence immunity is being lost in time (see Table \ref{table_1}). The first, second and third waves can be clearly seen at $t\approx 17000$, $t\approx 34000$ and $t\approx 50000$, respectively. As the scale of the colour bar is logarithmic, in the first wave of infections depicted by red in the spatiotemporal plot, a significant fraction of the population, between $1\%$ and $10\%$, has been infected. Similarly for the second and third peaks, leaving the community at an infected status, which is however lower than the ``Bronze'' community.

In the case of the ``Gold'' community, since the percentage of the vaccinated population is very high (i.e., $\phi_3=80\%$), the population still remains well immune even after a very long time compared with the infection levels in the ``Bronze'' and ``Silver'' communities.

The results in Fig. \ref{fig_three_non_connected_BSG} provide the baseline case for three unconnected ``Bronze'', ``Silver'' and ``Gold'' communities where the study of interconnected communities influencing each other will be based on in the following.

\subsection{Pair-wise interactions in ``Bronze''-``Gold" and ``Bronze''-``Silver" communities}\label{subsec_pair-wise_interactions_B_S_G_communities}

\begin{figure*}[!ht]
\centering
\includegraphics[height=18cm,width=0.99\textwidth]{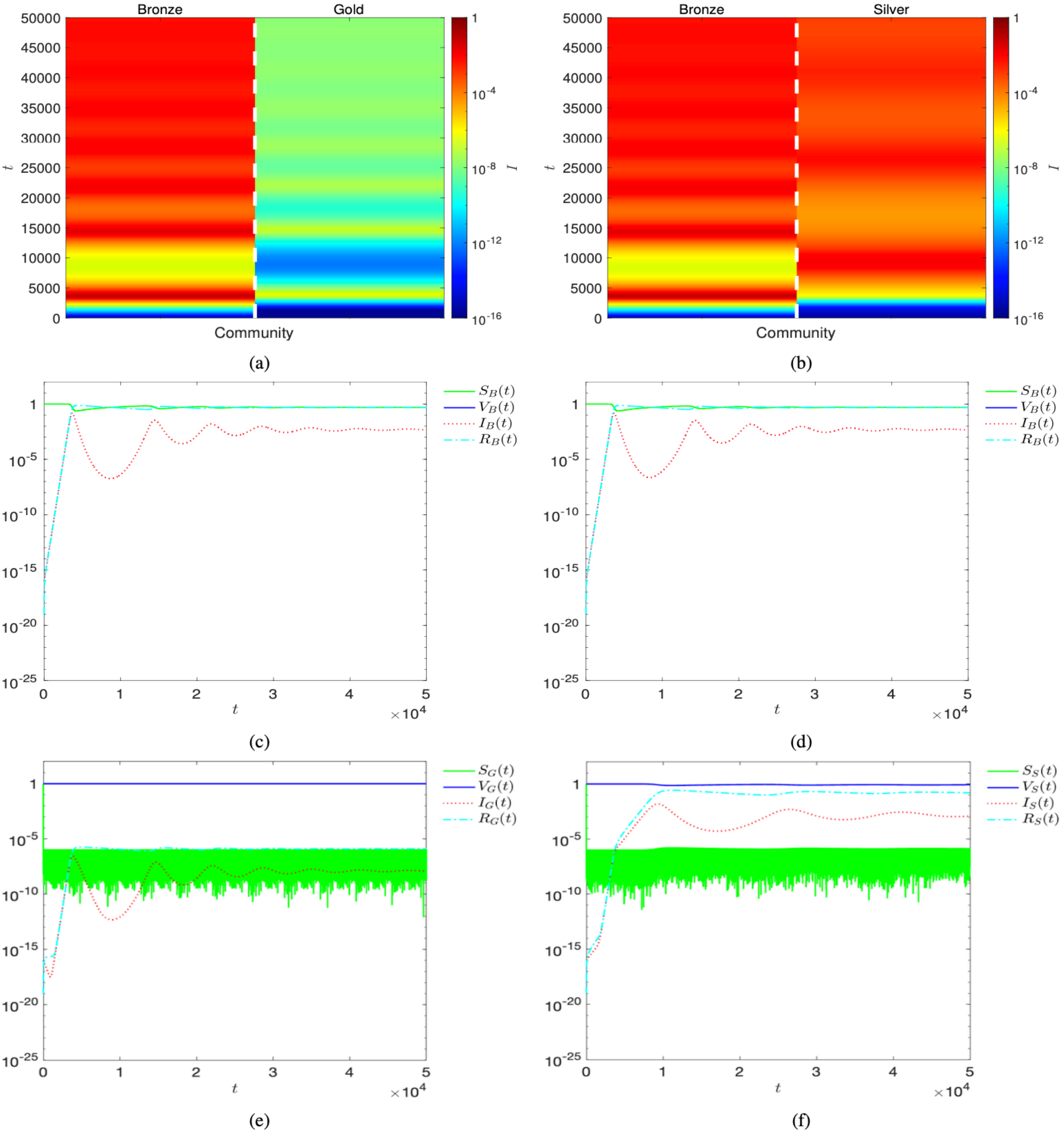}
 	\caption{Spatiotemporal plots of the evolution of infections, $I$, and evolution of $S$, $V$, $I$ and $R$ in time for a pair of ``Bronze''-``Gold" and ``Bronze''-``Silver" communities modelled by system \eqref{SVIR_coupled_system_ODEs}, where $\alpha=10^{-8}$. The horizontal axes in panels (a) and (b) denote the community index and the vertical the time $t$. In all other panels, the horizontal axes are linear and denote the time $t$ and the vertical are logarithmic and denote $S$, $V$, $I$ and $R$ of the ``Bronze'', ``Silver'' and ``Gold'' communities. The colour bars encode the infections $I_i,\;i=1,2$ in a logarithmic scale to depict small variations in infections near 0 as $I$ takes values in $[0,1]$. The vertical white, dashed lines in panels (a) and (b) delineate the two communities. Panels (a) and (b) are the spatiotemporal plots of infections for the pairs of ``Bronze''-``Gold" and ``Bronze''-``Silver" communities, respectively. Panels (c) and (e) show the evolution in time of $S$, $V$, $I$ and $R$ of the ``Bronze'' and ``Gold" communities in panel (a). Similarly, panels (d) and (f) show the evolution in time of $S$, $V$, $I$ and $R$ of the ``Bronze'' and ``Silver" communities in panel (b). Note that the subscript $B$, $S$ and $G$ in $S$, $V$, $I$ and $R$ denote ``Bronze'', ``Silver'' and ``Gold'', respectively.}\label{fig_two_communities}
\end{figure*}

Next we consider the simplest case of pair-wise interactions, studying the influence in infections in the pair of interconnected ``Bronze'' and ``Gold'' and ``Bronze'' and ``Silver'' communities. We want to study the influence in infections of the unvaccinated ``Bronze'' community to the very-well-vaccinated ``Gold'' community (``Bronze''-``Gold") and to a community vaccinated to a lesser degree, i.e., to the ``Silver'' community. We present the results of this analysis in Fig. \ref{fig_two_communities}, where we have set $\alpha=10^{-8}$ in system \eqref{SVIR_coupled_system_ODEs} to account for a small coupling strength in the connectivity of the pair of communities.

In particular, panel (a) shows the spatiotemporal plot of infections for a pair of ``Bronze'' and ``Gold'' communities, whereas panel (b) the spatiotemporal plot for a pair of ``Bronze'' and ``Silver'' communities. We observe that the evolution of infections in the ``Bronze'' communities in both panels is very similar and that first and subsequent infection waves occur more or less at the same time. This means that the dynamics of infections in the two ``Bronze'' communities remains unaltered when they come in contact with a ``Gold'' or ``Silver'' community and that moreover, the ``Gold'' and ``Silver'' communities cannot reduce the number of infections in the ``Bronze'' community. To the contrary, the ``Bronze'' community, which is unvaccinated, is driving the increase in infections in the ``Gold'' and ``Silver'' communities, even though these are significantly more immune than the ``Bronze'' community. Hence the ``Bronze'' community does not get affected when it comes in contact with the ``Gold'' and ``Silver'' communities. Indeed, while an isolated ``Gold'' community is in a steady healthy state as it can be seen in Fig. \ref{fig_three_non_connected_BSG}, once it comes in contact with the unvaccinated ``Bronze'' community, encounters the same first and secondary waves as the ``Bronze'' community, with less though intensity.

The total number of populations $S$, $V$, $I$, and $R$ of each community is an important feature of system \eqref{SVIR_coupled_system_ODEs}. To understand if individual communities or the system approaches an equilibrium point due to the continuous exchange between them, we plot in panels (c), (e) in Fig. \ref{fig_two_communities} for the pair of ``Bronze'' and ``Gold'' communities and in panels (d), (f) of the same figure for the pair of ``Bronze'' and ``Silver'' communities the evolution of $S$, $V$, $I$ and $R$ in time. Since the coupling strength here is very small, (i.e., $\alpha=10^{-8}$), one reasonably expects the equilibrium points of system \eqref{SVIR_coupled_system_ODEs} to be very close to those of system \eqref{svir_uncoupled} that we discuss in Subsec. \ref{subsec_esa_single_SVIR_model}. In the case of the pair of ``Bronze'' and ``Gold'' communities, $S$, $V$, $I$, $R$ of the ``Bronze'' community approach 0.28, 0, 0.0004, 0.72, respectively at the end of the integration, meaning that the ``Bronze'' community does not converge to the stable (as $\rho\beta-\lambda<1$) disease-free equilibrium point $E_0$, with the values showing that most of the population becomes recovered, to a less extend susceptible and to some extend infected. However, the ``Gold'' community approaches the stable disease-free equilibrium point $E_0$ as its $S$, $V$, $I$, $R$ values approach 0, 1, $10^{-8}$, $2\;10^{-6}$, respectively at the end of the integration. Comparing the values of the infected populations for the ``Bronze'' and ``Gold'' communities, we see that as expected from panel (a), the infected population in the ``Bronze'' community is much higher than in the ``Gold'' community at the end of the integration. Following a similar analysis for the pair of ``Bronze'' and ``Silver'' communities in panels (e) and (f) in Fig. \ref{fig_two_communities}, we have confirmed (running also longer simulations that are not shown here) that the ``Bronze'' community behaves the same way as in the case of the ``Bronze'' and ``Gold'' communities and that $S$, $V$, $I$, $R$ of the ``Silver'' community for intermediate times as in panels (b) and (f) approach the unstable disease-free equilibrium $E_0$ (see panel (f)), before they get attracted for longer integration times by the stable equilibrium \eqref{eq_fp_R0=1.2}. Finally, comparing the values of the infected populations for the ``Bronze'' and ``Silver'' communities, we see that as expected from panel (b), the infected populations in the ``Bronze'' and ``Silver'' communities are similar at the end of the integration.

Concluding, in these pairwise interactions, the unvaccinated ``Bronze'' community is driving the increase in infections in better vaccinated communities, such as the ``Gold'' and ``Silver'' communities, which shows the detrimental one-way effect of the ``Bronze'' to the ``Gold'' and ``Silver'' communities.

Next, we move on studying the dynamics in infections in a group of a ``Gold'', ``Silver'' and ``Bronze'' communities.

\subsection{Tripartite interactions in ``Bronze''-``Silver''-``Gold'' communities}\label{subsec_tripartite_interactions_B_S_G_communities}

\begin{figure*}[!ht]
\centering
\includegraphics[height=18cm,width=0.99\textwidth]{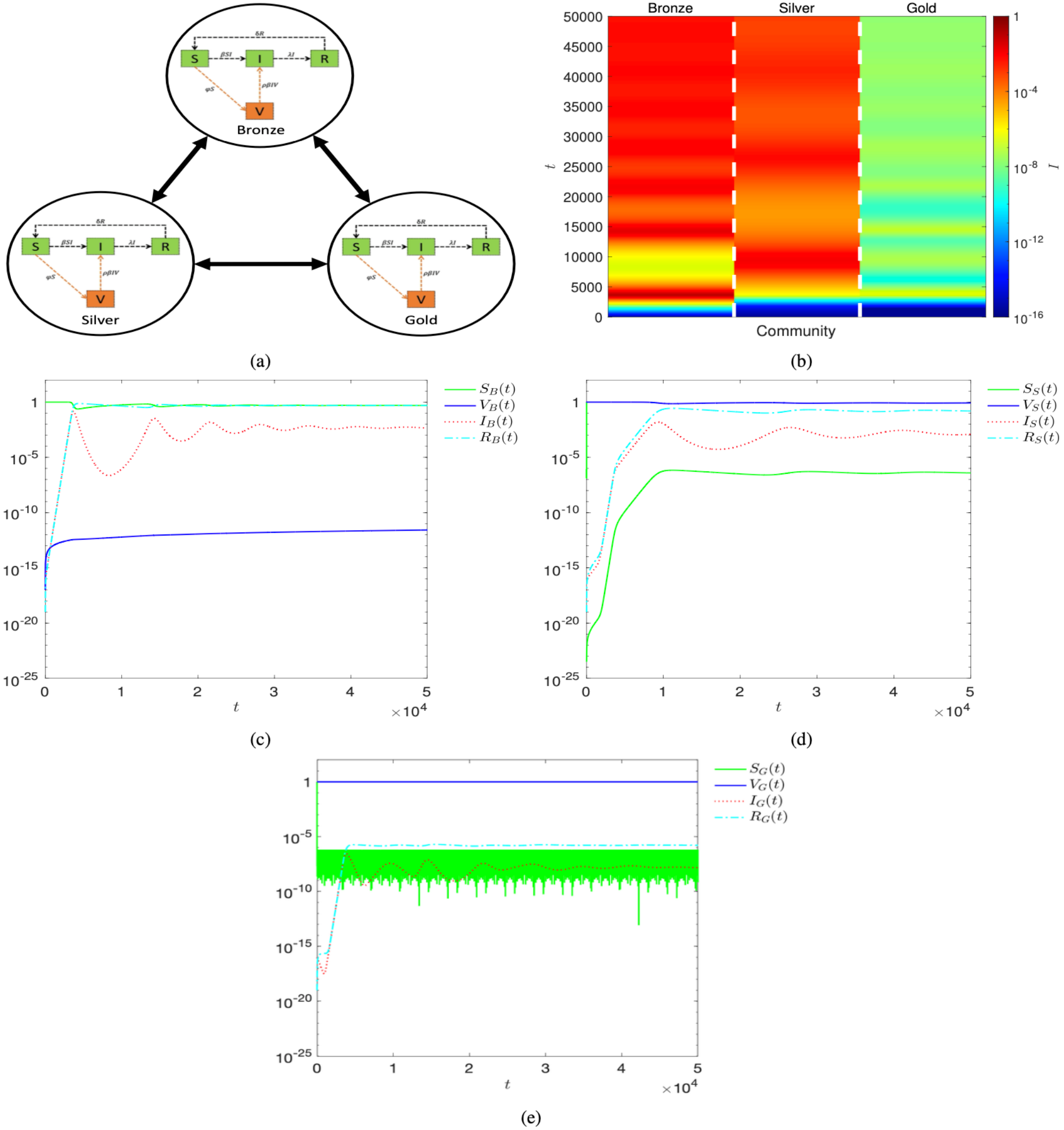}
%
%
\caption{All-to-all connectivity, spatiotemporal plot for three interconnected ``Bronze'', ``Silver'' and ``Gold'' communities and evolution in time of $S$, $V$, $I$ and $R$ for the same three communities, modelled by system \eqref{SVIR_coupled_system_ODEs}, where $\alpha=10^{-8}$. Panel (a) shows the three types of communities being connected in an all-to-all fashion and panel (b) the spatiotemporal plot of infections for the same three communities. Panels (c), (d) and (e) show the time evolution of $S$, $V$, $I$ and $R$ of the ``Bronze'', ``Silver'' and ``Gold'' communities, respectively. The horizontal axes in panel (b) denote the community index and the vertical the time $t$. The horizontal axes in panels (c), (d) and (e) are linear and denote the time $t$ and the vertical logarithmic and denote $S$, $V$, $I$ and $R$ of the ``Bronze'', ``Silver'' and ``Gold'' communities. The colour bar in panel (b) encodes the infections $I_i,\;i=1,2,3$ in a logarithmic scale to depict small variations in infections near 0 as $I$ takes values in $[0,1]$. In panel (b), the vertical white, dashed lines delineate the three communities. Note that the subscript $B$, $S$ and $G$ in $S$, $V$, $I$ and $R$ denote ``Bronze'', ``Silver'' and ``Gold'', respectively.}\label{fig_three_connected_BSG}
\end{figure*}

Here we study the dynamics of infections in a group of interconnected ``Bronze'', ``Silver'' and ``Gold'' communities. This can be thought of as a minimal fully connected network that contains all types of communities, as shown in panel (a) in Fig. \ref{fig_three_connected_BSG}.

Figure \ref{fig_three_connected_BSG}(a) shows that the three types of communities are connected in an all-to-all fashion. Comparing the spatiotemporal plot in Fig. \ref{fig_two_communities}(a) for a pair of connected ``Bronze'' and ``Gold'' communities with the plot in Fig. \ref{fig_three_connected_BSG}(b) for the trio of ``Bronze'', ``Silver'' and ``Gold'' communities, we see that the dynamics of infections in the ``Gold'' community does not influence any of the infection dynamics of the ``Bronze'' and ``Silver'' communities. Hence the influence of the ``Bronze'' community to the infections in the ``Gold'' and ``Silver'' communities is noticeable. Even though the ``Silver'' community is better vaccinated than the ``Bronze'', it is not able to stop the ``Bronze'' community driving the rise in infections in the ``Gold'' community. Where we had a reduction in the number of infections in the ``Gold'' community in Fig. \ref{fig_two_communities}(a), depicted as the blue strip between 5000 and 10000 time, we have an increase in infection levels in Fig. \ref{fig_three_connected_BSG}(b). Actually, this comparison reveals that the successive waves are not only shifted in time but also become intensified.

As in the case of the pairs of ``Bronze''-``Gold" and ``Bronze''-``Silver" communities in Subec. \ref{subsec_pair-wise_interactions_B_S_G_communities}, the total number of populations $S$, $V$, $I$, and $R$ for each community is an important feature of system \eqref{SVIR_coupled_system_ODEs}. In particular, we want to understand in the case of tripartite interactions between ``Bronze'', ``Silver'' and ``Gold'' communities, if individual communities or the system approaches an equilibrium point due to the continuous exchange between them. To this end, we plot in panels (c), (d) and (e) in Fig. \ref{fig_three_connected_BSG} the evolution of $S$, $V$, $I$ and $R$ in time of the three interconnected communities. We see that as in the case of ``Bronze''-``Gold" and ``Bronze''-``Silver'' community interactions in Subec. \ref{subsec_pair-wise_interactions_B_S_G_communities}, the ``Bronze'' community does not converge to the disease-free equilibrium point $E_0$, but evolves around values that show that most of the population becomes recovered, to a less extend susceptible and to some extend infected. The ``Silver'' community for intermediate times as in panels (b) and (c), approach the unstable disease-free equilibrium $E_0$ (see panel (d)), before they get attracted for longer integration times by the stable equilibrium \eqref{eq_fp_R0=1.2} (plot not shown here). The ``Gold'' community approaches the stable disease-free equilibrium point $E_0$ in panel (e), where almost the whole population becomes vaccinated at the end of the integration. Finally, comparing the values of the infected populations for the three communities, we see that as expected from panel (b), the infected populations in the ``Bronze'' and ``Silver'' communities are similar and those of the ``Gold'' community music smaller at the end of the integration.

\subsection{Spread of a virus in a BA scale-free network}\label{subsec_spread_virus_in_a_BA_network}

We now move on to the study of the spread of a virus in a larger-size network given by a BA scale-free graph. To this end, we use the network of $N=60$ nodes in panel (c) in Fig. \ref{fig_ERN_WSN_BN_60_nodes} with network density 0.033. Here we study the spread of a virus in heterogeneous (with regard to $\beta,\rho,\phi,\lambda,\delta$) communities connected in a network with a few hubs. A commonly found feature in real world networks is the presence of hubs, which are nodes that are highly connected to other nodes in the network. This means that hubs are nodes with a much higher node-degree than most other nodes in the network. This has important consequences for how information, viruses or diseases travel or spread in the network. Such networks are scale-free as they are characterised by a power-law degree distribution. Several natural and human-made systems, e.g., social networks can be approximated by scale-free networks as they contain few nodes with high degree as compared to other nodes in the network. The BA model \cite{RevModPhys.74.47} is an algorithm for generating scale-free networks using a preferential attachment mechanism, meaning that the more connected a node is, the more likely it is to receive new connections. Many observed networks fall approximately into the class of scale-free networks, meaning that they have power-law degree distributions, while ER random and WS small-world networks do not exhibit power laws. The BA network used here and its adjacency or connectivity matrix $A^I$ was computed using the igraph package in R. Then, the Laplacian matrix $L^I$ was computed from $A^I$, using Eq. \eqref{laplacian_matrix}, which was then fed into system \eqref{SVIR_coupled_system_ODEs} to integrate it numerically.

In particular, we want to study the effect of unvaccinated (i.e., ``Bronze'') communities to the infection levels of moderate (i.e., ``Silver'') and well-vaccinated (i.e., ``Gold'') communities in time for weak coupling, e.g. for $\alpha=10^{-8}$. To this extend, we are studying four distinct cases of community configurations, namely (A) where the first community is ``Bronze'', the next 29 are ``Silver'' and the last 30 are ``Gold'', (B) the first 5 are ``Bronze'', the next 25 are ``Silver'' and the last 30 are ``Gold'', (C) the first 20 are ``Bronze'', the next 20 ``Silver'' and the last 20 ``Gold'' and (D) the first 40 are ``Bronze'', the next 19 ``Silver'' and the last one is ``Gold''.

\begin{figure*}[!ht]
\centering
 	\includegraphics[height=17.1cm,width=0.99\textwidth]{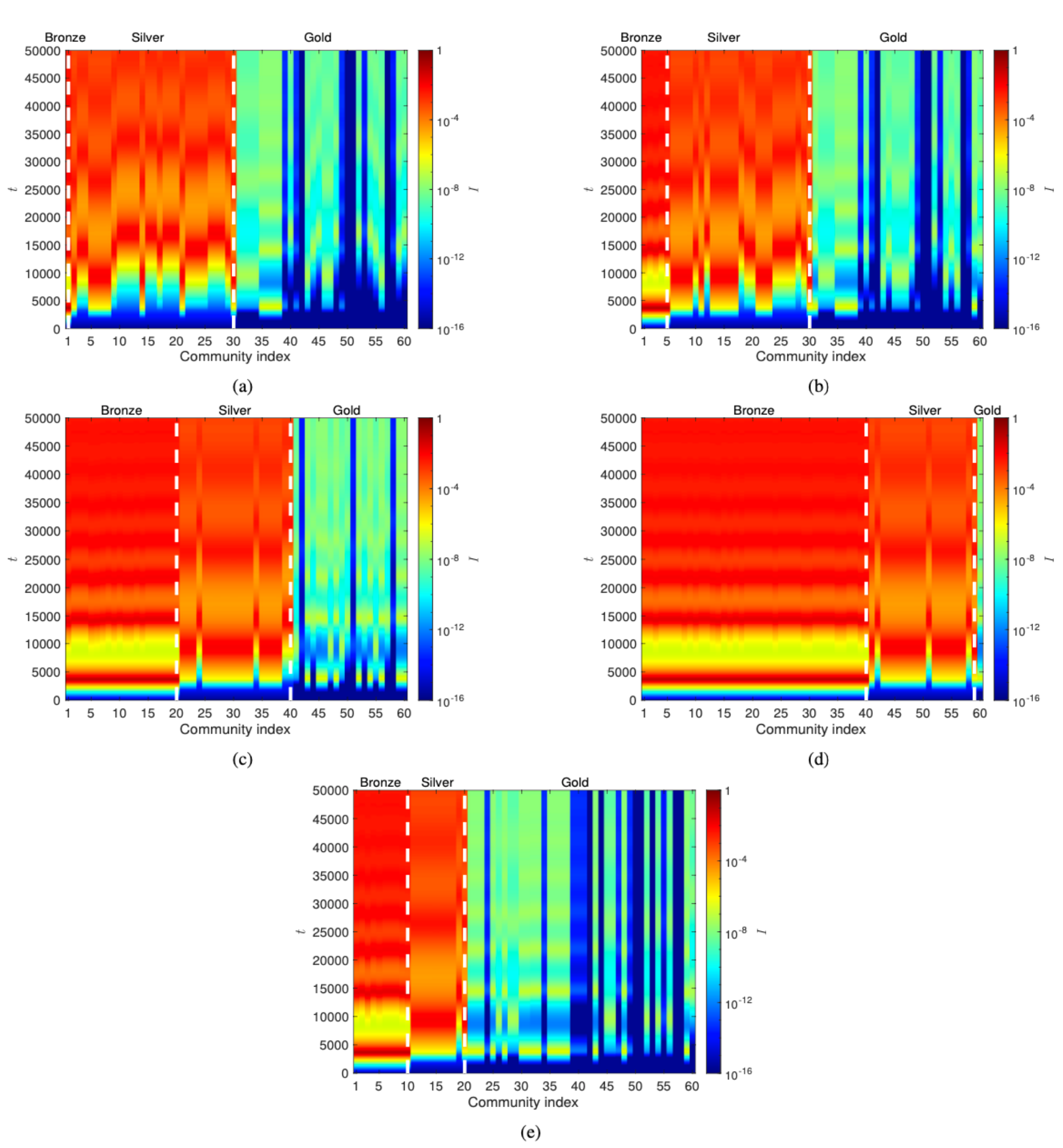}
\caption{Spatiotemporal plots of infections for the community configurations (A), (B), (C), (D) and (E) discussed in Subsec. \ref{subsec_spread_virus_in_a_BA_network}, where $\alpha=10^{-8}$. Panel (a) shows the spatiotemporal plot of infections in case (A) of 1 ``Bronze'', 29 ``Silver'' and 30 ``Gold'' communities. Panel (b) is for case (B) of 5 ``Bronze'', 25 ``Silver'' and 30 ``Gold"" communities. Panel (c) shows the spatiotemporal plot of infections for case (C) of 20 ``Bronze'', 20 ``Silver'' and 20 ``Gold'' communities. Panel (d) shows the spatiotemporal plot of infections for case (C) of 40 ``Bronze'', 19 ``Silver'' and 1 ``Gold'' communities. Panel (e) shows the spatiotemporal plot of infections for case (C) of 10 ``Bronze'', 10 ``Silver'' and 40 ``Gold'' communities. The communities are connected via the BA network of $N=60$ nodes in Fig. \ref{fig_ERN_WSN_BN_60_nodes}(c). The horizontal axes denote the community index and the vertical the time $t$. The colour bar encodes the infections $I_i,\;i=1,\ldots,60$ in a logarithmic scale to depict small variations in infections near 0 as $I$ takes values in the interval $[0,1]$. The vertical white, dashed lines delineate the ranges of ``Bronze'', ``Silver'' and ``Gold'' communities.}\label{fig_BN_different_community_configurations_60_nodes}
\end{figure*}

We report on the results of this analysis in Fig. \ref{fig_BN_different_community_configurations_60_nodes}, where panel (a) shows the trajectory of infections in case (A), where the first community is ``Bronze'', the next 29 are ``Silver'' and the last 30 are ``Gold''. One can see that the trajectory of infections in the ``Bronze'' community is very similar to the infection trajectory in the ``Bronze'' community when it is not connected to the ``Silver'' and ``Gold'' communities (see Fig. \ref{fig_three_non_connected_BSG}) and of the ``Bronze'' community in the trio of weakly connected (i.e., $\alpha=10^{-8}$) ``Bronze'', ``Silver'' and ``Gold'' communities in Fig. \ref{fig_three_connected_BSG}(b). Moreover, one also observes similar secondary waves of infections springing out in time even though there are no infected individuals initially due to the interactions of the unvaccinated ``Bronze'' community with the ``Silver and ``Gold'' communities. Since the ``Bronze'' community is not vaccinated, its infections reaches 1 after long time depicted by red, meaning that almost all of the population in the ``Bronze'' community becomes infected. Interestingly, even though the ``Bronze'' community interacts with the ``Silver'' and ``Gold'' communities, which are better vaccinated, the latter are not able to reduce the infection levels in the ``Bronze'' community, as the ``Bronze'' community is unvaccinated. The fate of infections in the ``Silver'' and ``Gold'' communities is similar again to those in Fig. \ref{fig_three_non_connected_BSG}) for the unconnected trio of communities and to those in \ref{fig_three_connected_BSG}(b) for the weakly connected trio of ``Bronze'', ``Silver'' and ``Gold'' communities. Our analysis shows that the infections in the moderately vaccinated ``Silver'' and well-vaccinated ``Gold'' communities are much lower than those in the unvaccinated ``Bronze'' community, hence the latter are able to resist the spread of the virus. When such types of equally-sized types of communities interact weakly with each other, then the infections remain at the same levels as when they are not interacting and the well-vaccinated ``Gold'' communities are the best in keeping their infection levels at bay.

This does not seem to change even when the number of unvaccinated ``Bronze'' communities increases in panels (b), (c) and (d) in Fig. \ref{fig_BN_different_community_configurations_60_nodes} from 5 in (b) to 20 in (c) to 40 in (d). Still the same conclusions can be drawn and the ``Gold'' communities are the best in keeping the infections at bay, even when they come into contact with less vaccinated (i.e., ``Silver'') or unvaccinated (i.e., ``Bronze'') communities. Another important conclusion is that looking at panels (a), (b), (c) and (d) in Fig. \ref{fig_BN_different_community_configurations_60_nodes}, only highly vaccinated communities are able to resist the spread of the virus as evidently, the infections in the moderately vaccinated ``Silver'' communities soar at the same levels and share the same time-scales of the ``Bronze'' communities when they interact. Even in the well-vaccinated ``Gold'' communities, the infections increase by some orders of magnitudes, they are however much lower than those in the ``Bronze'' and ``Silver'' communities.

In Fig. \ref{fig_BN_different_community_configurations_60_nodes}(d) we study the interesting case where most of the communities are ``Gold'' (i.e., very well-vaccinated) and the effect they have to ``Silver'' and ``Bronze'' communities, which are less or not vaccinated at all. It is evident that even though the majority of the communities are ``Gold'', they cannot help reduce the infection levels in the ``Bronze'' and ``Silver'' communities, where infections soar to similar levels as in panels (a) to (d).

Concluding, even very few unvaccinated ``Bronze'' communities are enough to lead to increase in infection levels in moderately and well-vaccinated weakly connected communities, perturbing once and for all the very low levels of initial infections in ``Gold'' and ``Silver'' communities. Even a large number of ``Gold'' communities is not enough to revert the levels of infections and help reduce them.

\subsection{Statistical analysis of the spread of a virus in ER random, WS small-world and BA scale-free networks}\label{subsec_stats_analysis}

In the last section, we studied the spread of a virus in a single, sparse, BA scale-free network for different numbers of ``Bronze'', ``Silver'' and ``Gold'' communities, running single simulations of model \eqref{SVIR_coupled_system_ODEs} in each case. Here we expand our study to ER random \cite{Erdosetal1959}, WS small-world \cite{Wattsetal1998} and BA scale-free \cite{RevModPhys.74.47} networks of $N=60$ communities, performing a statistical analysis and reporting on results over a sample of 100 different ER random, WS small-world and BA scale-free networks. We focus on case (B) in Sec. \ref{subsec_spread_virus_in_a_BA_network} and consider that the first 5 communities are ``Bronze'', the next 25 are ``Silver'' and the last 30 are ``Gold''. In the case of a single BA scale-free network, we have found that even a small number of unvaccinated ``Bronze'' communities is enough to lead to an increase in infection levels in moderately and well-vaccinated weakly connected (i.e., for $\alpha=10^{-8}$) communities, perturbing once and for all the very low levels of initial infections in ``Gold'' and ``Silver'' communities. Here, we revisit the same question and want to see if a small number of unvaccinated ``Bronze'' communities can affect the spread of the virus in moderately vaccinated ``Silver'' and well-vaccinated ``Gold'' communities or whether well-vaccinated communities when connected to moderately or non-vaccinated communities can help reduce their infection levels in ER random, WS small-world and BA scale-free networks. Again the  ER random networks we have used in this analysis have rewiring probability 0.3 (resulting in network densities 0.3), the WS small-world networks network density 0.07 and the BA scale-free networks densities around 0.033. We also want to find if the results are robust when addressing the question for a sample of such types of networks and if network topology plays a role in the spread of the virus and in infection levels. In our analysis, we have opted for 100 different ER random, WS small-world and BA scale-free networks and report on average and standard deviations over the 100 networks. The idea behind using these types of networks is that they share different structural  properties, found also in real-life networks \cite{Newman2010}.

\begin{figure*}[!ht]
\centering
\includegraphics[height=17.1cm,width=0.99\textwidth]{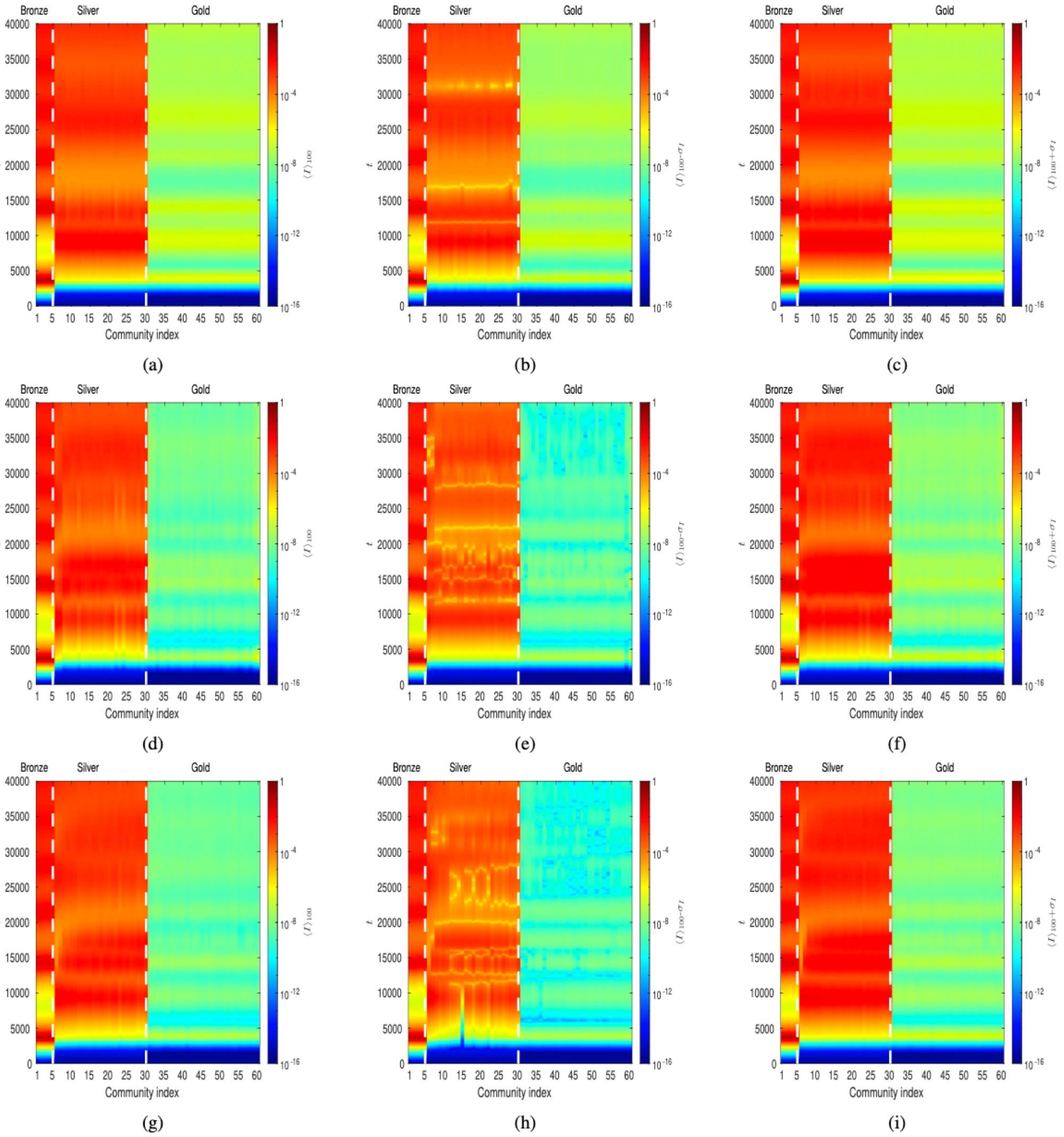}
\caption{Spatiotemporal plots of infection levels for 100 ER random, 100 WS small-world and 100 BA scale-free networks for weak connectivity (i.e., for $\alpha=10^{-8}$). Panels (a) to (c) show the spatiotemporal plot of infection levels averaged over 100 ER random networks. Panels (d) to (f) similarly for 100 WS small-world networks. Panels (g) to (i) for 100 BA scale-free networks. The panels in the left column present the average infection levels over the 100 networks, $\langle I\rangle_{100}$, and the panels in the middle and right columns the average infection levels minus one standard deviation, $\langle I\rangle_{100}-\sigma_I$, and average infection levels over the 100 networks plus one standard deviation over the 100 networks, $\langle I\rangle_{100}+\sigma_I$, respectively. The horizontal axes denote the community index and the vertical the time $t$. The scale of the colour bar is logarithmic to depict small variations in infections near 0 as infections $I$ take values in the interval $[0,1]$. The vertical white, dashed lines delineate the ranges of ``Bronze'', ``Silver'' and ``Gold'' communities.}\label{fig_stats_ERN_SWN_BN}
\end{figure*}

We summarise the results of our statistical analysis for weakly connected communities (i.e., for $\alpha=10^{-8}$) in Fig. \ref{fig_stats_ERN_SWN_BN} where we present the spatiotemporal plots of infection levels over time for 100 ER random networks in panels (a) to (c), for 100 WS small-world networks in panels (d) to (f) and for 100 BA scale-free networks in panels (g) to (i). The panels in the left column present the average infection levels over the 100 networks, $\langle I\rangle_{100}$, and the panels in the middle and right columns, the average infection levels minus one standard deviation, $\langle I\rangle_{100}-\sigma_I$, and average infection levels over the 100 networks plus one standard deviation over the 100 networks, $\langle I\rangle_{100}+\sigma_I$, respectively. Our results show that the infection levels in WS small-world and BA scale-free networks are a little bit lower to those in ER random networks but not as big to make a difference, both in the short and long term. However, for all types of networks, 5 unvaccinated ``Bronze'' communities are enough to lead to an increase in infection levels in moderately (``Silver'') and well-vaccinated (``Gold'') weakly connected communities, increasing once and for all the very low levels of initial infections in ``Gold'' and ``Silver'' communities. This is because the initial horizontal blue stripes (of very low infection levels) in all cases of networks turn quickly to a series of yellow and red (emergence of secondary surges or waves) and eventually to red of very high levels of infections in all types of networks. Another reason is that our model \eqref{SVIR_coupled_system_ODEs} assumes there is a constant rate for the loss of immunity $\delta=0.0001$ (with $1/\delta$ being the mean immune period \cite{VARGASDELEON20111106}), in other words that the effect of the vaccine wanes in time and that the vaccination has efficacy $0\leq \rho\leq 1$, different for each type of community, given in Table \ref{table_1}. This shows that actually the trajectory of infection levels in interconnected communities is not influenced by the network topology, but by the unvaccinated ``Bronze'' communities, as these are responsible for the ``Silver'' and ``Gold'' communities to become highly infected at the end. Even a small number of ``Bronze'' communities is enough to drive the infection levels in moderately and well-vaccinated communities to an increase, regardless of the type of network.

\subsection{Spread of a virus in networks of different densities, coupling strengths and inhomogeneous communities}\label{subsec_strong_coupling}

\begin{figure*}[!ht]
	\centering
\includegraphics[height=5.7cm,width=0.99\textwidth]{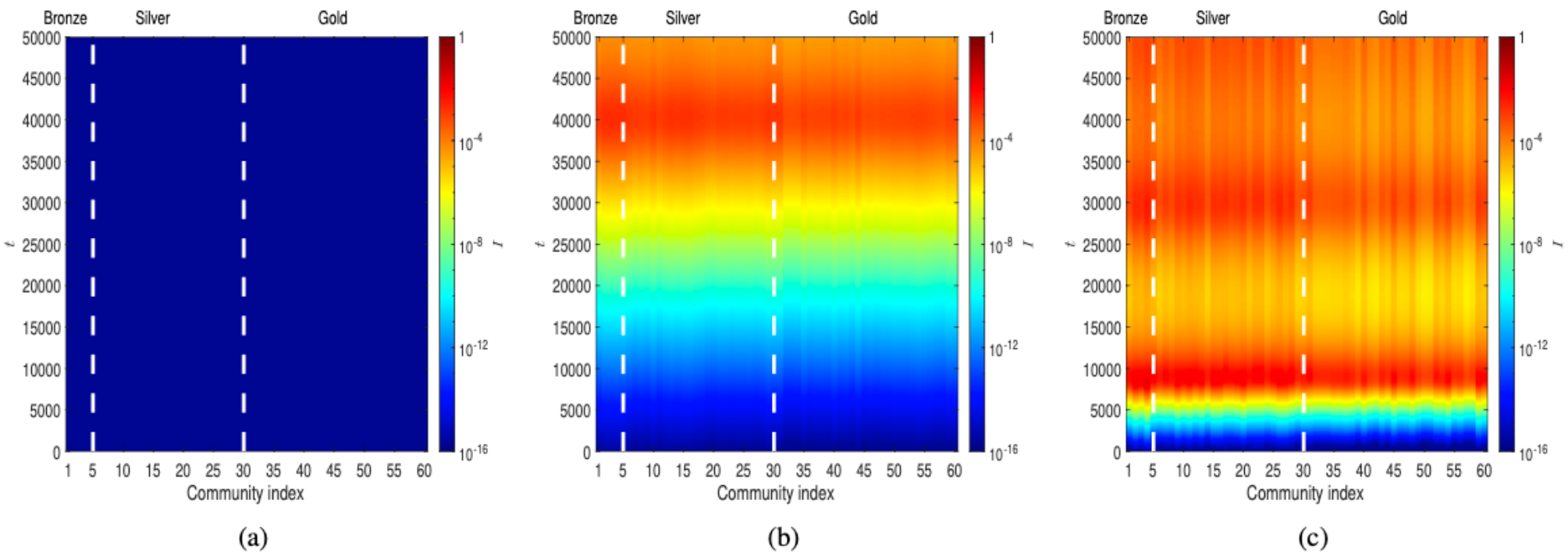}
\caption{Spatiotemporal plots of infections for 5 ``Bronze'', 25 ``Silver'' and 30 ``Gold'' connected communities by means of ER random, WS small-world and BA scale-free networks, where $\alpha=0.01$ in system \eqref{SVIR_coupled_system_ODEs}. Panel (a) shows the spatiotemporal plot of infections for an ER random network. Panel (b) is similar to panel (a) for a WS small-world network. Panel (c) is similar to the other two panels for a BA scale-free network. The horizontal axes denote the community index and the vertical the time $t$. The colour bar encodes the infections $I_i,\;i=1,\ldots,60$ in a logarithmic scale to depict small variations in infections near 0 as $I$ takes values in the interval $[0,1]$. The vertical white, dashed lines delineate the ranges of ``Bronze'', ``Silver'' and ``Gold'' communities.}\label{fig_a2=0.01_thermalisation}
\end{figure*}

Finally, here we address the question of the trajectory of infection levels in ER random, WS small-world and BA scale-free networks with different network densities, coupling strengths and inhomogeneous communities. In the context of model \eqref{SVIR_coupled_system_ODEs}, strong connectivity could be interpreted as the easiness with which infected individuals from ``Bronze'', ``Silver'' and ``Gold'' communities travel among them spreading the virus. This is modelling for example how easy it is for infected individuals to travel from one country to another. To this end, we study the same question in Subsec. \ref{subsec_spread_virus_in_a_BA_network}, focusing again on the paradigmatic case (B), where in a network of $N=60$ communities, the first 5 are ``Bronze'', the next 25 ``Silver'' and the last 30 ``Gold''. In the previous sections, we have found that a small number of unvaccinated ``Bronze'' communities are enough to lead to increased infection levels in moderately and well-vaccinated weakly connected (i.e., for $\alpha=10^{-8}$) communities with the ``Gold'' ones being the most successful in keeping the increase in infection levels at bay as they are the best vaccinated with the best vaccine efficacy. Here, we revisit the same question in Fig. \ref{fig_a2=0.01_thermalisation} considering the ER random, WS small-world and BA scale-free networks shown in Fig. \ref{fig_ERN_WSN_BN_60_nodes} with densities 0.3, 0.07 and 0.0033, respectively and in Fig. \ref{fig_sec_network_properties_and_the_spread_of_a_virus} the same types of networks where they all have the same network density 0.3. We also consider the coupling strengths $\alpha=0.01$ in Fig. \ref{fig_a2=0.01_thermalisation} and $10^{-8}$ and 0.01 in Fig. \ref{fig_sec_network_properties_and_the_spread_of_a_virus}. We want to see if a small number of unvaccinated ``Bronze'' communities can affect the spread of the virus in moderately vaccinated ``Silver'' and well-vaccinated ``Gold'' communities. We also assume in the simulations that all populations are initially susceptible.

Figure \ref{fig_a2=0.01_thermalisation} shows the spatiotemporal plots of infections for 5 ``Bronze'', 25 ``Silver'' and 30 ``Gold'' connected communities by means of the ER random network in Fig. \ref{fig_ERN_WSN_BN_60_nodes}(a)), shown in panel (a), the WS small-world network, seen in Fig. \ref{fig_ERN_WSN_BN_60_nodes}(b)), in panel (b) and the BA scale-free network seen in Fig. \ref{fig_ERN_WSN_BN_60_nodes}(c)), in panel (c), where $\alpha=0.01$ in system \eqref{SVIR_coupled_system_ODEs}. The spatiotemporal plot in panel (a) shows that, starting initially with all populations being susceptible, the virus does not spread in the ER random network of connected communities as all $I_i$ become as small as $10^{-16}$ in time. In this case, the system reaches the disease-free state. This is a striking result when compared to the spatiotemporal plots for the sparser WS small-world and BA scale-free networks in panels (b) and (c), respectively. In the latter two cases, the virus spreads to all communities in the long term, regardless of whether they are vaccinated or not, or the percentage and efficacy of the vaccination. It is worth mentioning comparing panels (b) with (c) that (A) it takes more time for the virus to spread in a WS small-world network (panel (b)) than in a BA scale-free network (panel (c)), presumably due to the presence of hubs (which act as mega-spreaders) in the latter network and that (B) the ``Gold'' communities are not able to slow-down the spread of the virus when they come in contact with the ``Bronze'' and ``Silver'' communities. In the long term, all types of communities in the WS small-world and BA scale-free networks will be infected due to the efficacy, $\rho$ which is not maximum and the rate for the loss of immunity $\delta=0.0001$ for all communities (i.e., the efficacy of the vaccination is waning over time). This is reminiscent of the real-life situation whereby regardless of whether a population is vaccinated or not, or the percentage and efficacy of the vaccination, if it is not vaccinated periodically, it will become infected in the long term when coming in contact with other infected populations.

\begin{figure*}[!ht]
	\centering
\includegraphics[height=11.4cm,width=0.99\textwidth]{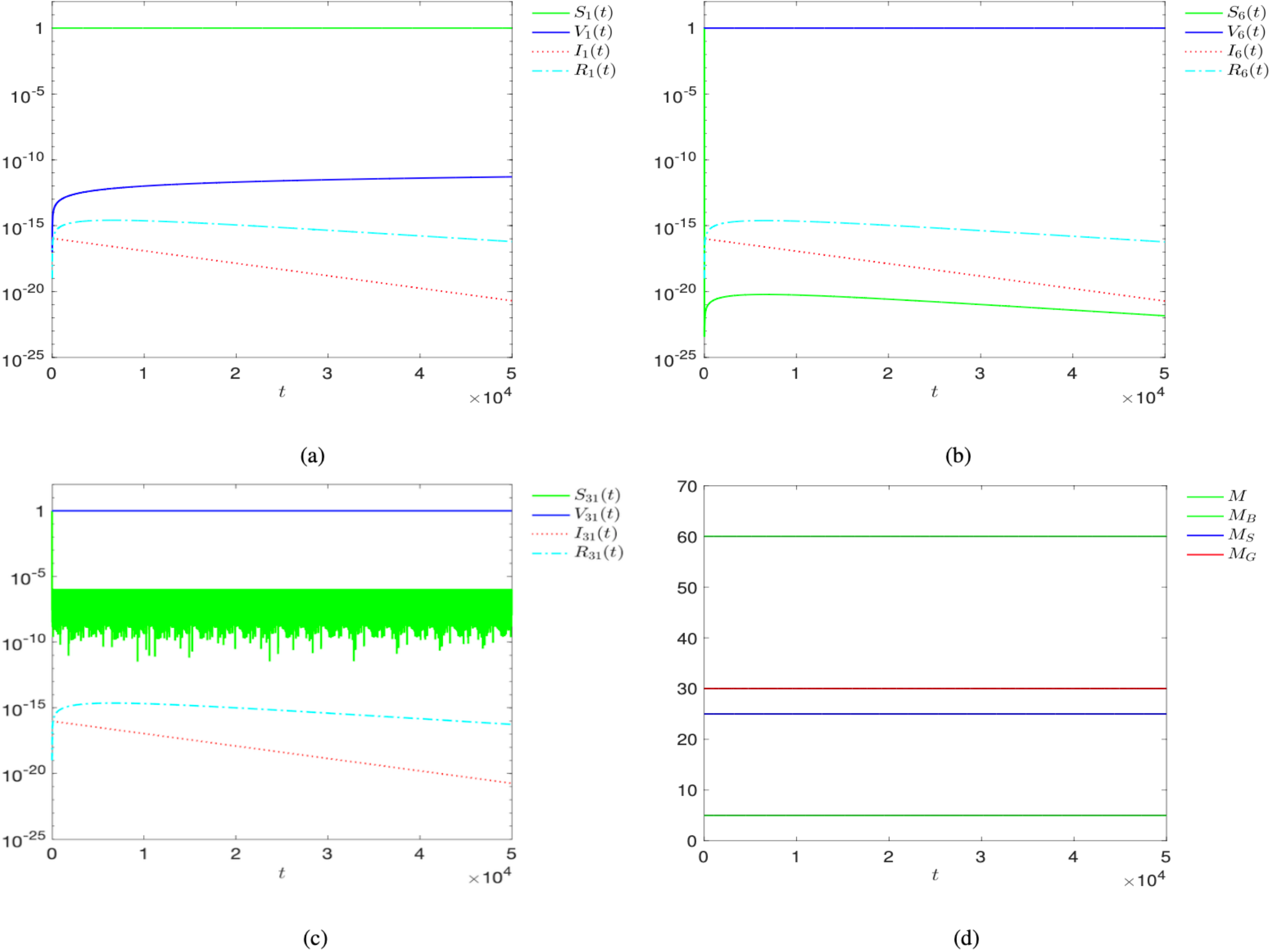}
\caption{Evolution of $S$, $V$, $I$, $R$, $M$, $M_B$, $M_S$ and $M_G$ in time for the configuration in panel (a) in Fig. \ref{fig_a2=0.01_thermalisation} for the ER random network, where $\alpha=0.01$ in system \eqref{SVIR_coupled_system_ODEs}. Panel (a) shows $S_1$, $V_1$, $I_1$, $R_1$ in time of the first ``Bronze'' community. Panel (b) shows $S_6$, $V_6$, $I_6$, $R_6$ of the first ``Silver'' community in time and panel (c) shows $S_{31}$, $V_{31}$, $I_{31}$, $R_{31}$ of the first ``Gold'' community in time. Panel (d) shows the evolution of $M$, $M_B$, $M_S$ and $M_G$ in time. The black lines in panel (d) correspond to the number of ``Bronze'', ``Silver'' and ``Gold'' communities and to the total population, which are 5, 25, 30 and 60, respectively.}\label{fig_a2=0.01_thermalisation_ERN}
\end{figure*}

In the following, we will elucidate the striking difference between the disease-free state in panel (a) for the ER random network and spread of the virus in panels (b) and (c) in WS small-world and BA scale-free networks, assuming that all populations are initially susceptible. To do so, we denote the populations of the ``Bronze'', ``Silver'' and ``Gold'' communities by $M_B$, $M_S$ and $M_G$, respectively, whereby the total constant population $M=M_B+M_S+M_G=N=60$. It is worth it to mention that in this framework, $M_B$, $M_S$ and $M_G$ are not necessarily constant in time, but can vary. In Fig. \ref{fig_a2=0.01_thermalisation_ERN}, we are showing the evolution of $S$, $V$, $I$, $R$, $M$, $M_B$, $M_S$ and $M_G$ in time for the configuration in panel (a) in Fig. \ref{fig_a2=0.01_thermalisation} for the ER random network, where $\alpha=0.01$ in system \eqref{SVIR_coupled_system_ODEs}. Panel (a) shows $S_1$, $V_1$, $I_1$, $R_1$ in time of the first ``Bronze'' community. Panel (b) shows $S_6$, $V_6$, $I_6$, $R_6$ of the first ``Silver'' community in time and panel (c) shows $S_{31}$, $V_{31}$, $I_{31}$, $R_{31}$ of the first ``Gold'' community in time. Panel (d) shows the evolution of $M$, $M_B$, $M_S$ and $M_G$ in time.

Panel (a) shows that in the ER random network, the first ``Bronze'' community becomes susceptible and panels (b), (c) that the first ``Silver'' and first ``Gold'' communities become vaccinated in the long term, assuming starting initially with susceptible populations. The same respective behaviours happen for all ``Bronze'', ``Silver'' and ``Gold'' communities. In the case of the ``Bronze'' communities, they become susceptible as $\phi_i=\rho_i=0$, $i=1,\ldots,5$, meaning that the second equation for all ``Bronze'' communities in system \eqref{SVIR_coupled_system_ODEs} is identically 0 at all times as the initial condition for $V_i$ is 0 for all $i=1,\ldots,5$. Panels (a) to (c) reveal that the infected populations of the first ``Bronze'', first ``Silver'' and first ``Gold'' communities reduce in time and become smaller than $10^{-20}$ at the end of the integration (i.e., at $t=5\times10^4$). We have checked that this is the case, respectively, for all ``Bronze'', ``Silver'' and ``Gold'' communities. This explains why the infected populations of all three types of communities are practically zero in panel (a) in Fig. \ref{fig_a2=0.01_thermalisation}. Panel (d) confirms that the populations of the ``Bronze'' ($M_B$), ``Silver'' ($M_S$) and ``Gold'' ($M_G$) communities remain constant in time as the total population $M$. Hence in the case of the ER random network, there are no subpopulations moving from one type of community to another in time and all communities become disease-free in the long term.

\begin{figure*}[!ht]
	\centering
\includegraphics[height=11.4cm,width=0.99\textwidth]{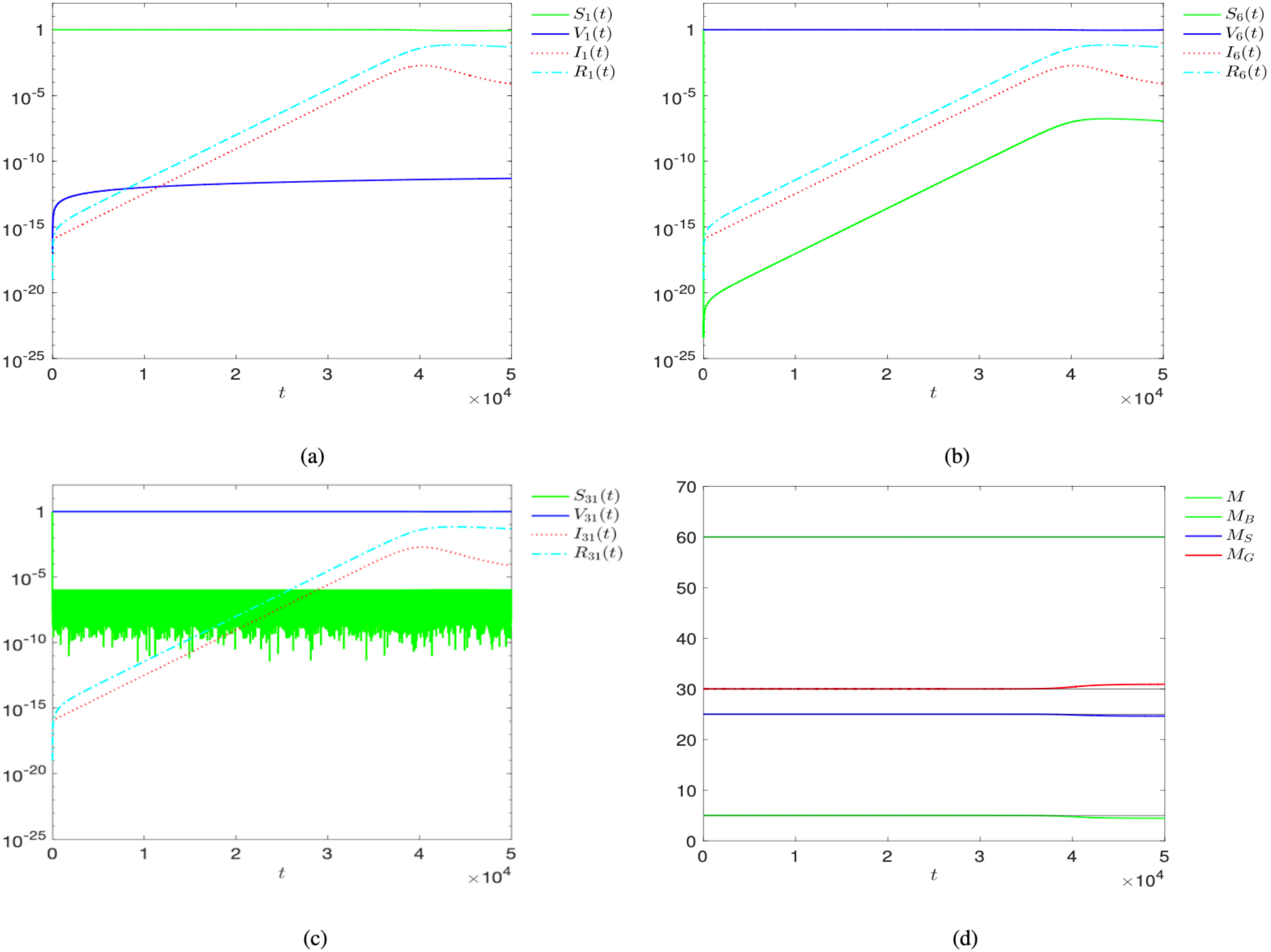}
\caption{Evolution of $S$, $V$, $I$, $R$, $M$, $M_B$, $M_S$ and $M_G$ in time for the configuration in panel (b) in Fig. \ref{fig_a2=0.01_thermalisation} for the WS small-world network, where $\alpha=0.01$ in system \eqref{SVIR_coupled_system_ODEs}. Panel (a) shows $S_1$, $V_1$, $I_1$, $R_1$ in time of the first ``Bronze'' community. Panel (b) shows $S_6$, $V_6$, $I_6$, $R_6$ of the first ``Silver'' community in time and panel (c) shows $S_{31}$, $V_{31}$, $I_{31}$, $R_{31}$ of the first ``Gold'' community in time. Panel (d) shows the evolution of $M$, $M_B$, $M_S$ and $M_G$ in time. The black lines in panel (d) correspond to the number of ``Bronze'', ``Silver'' and ``Gold'' communities and to the total population, which are 5, 25, 30 and 60, respectively.}\label{fig_a2=0.01_thermalisation_WS}
\end{figure*}

\begin{figure*}[!ht]
	\centering
\includegraphics[height=11.4cm,width=0.99\textwidth]{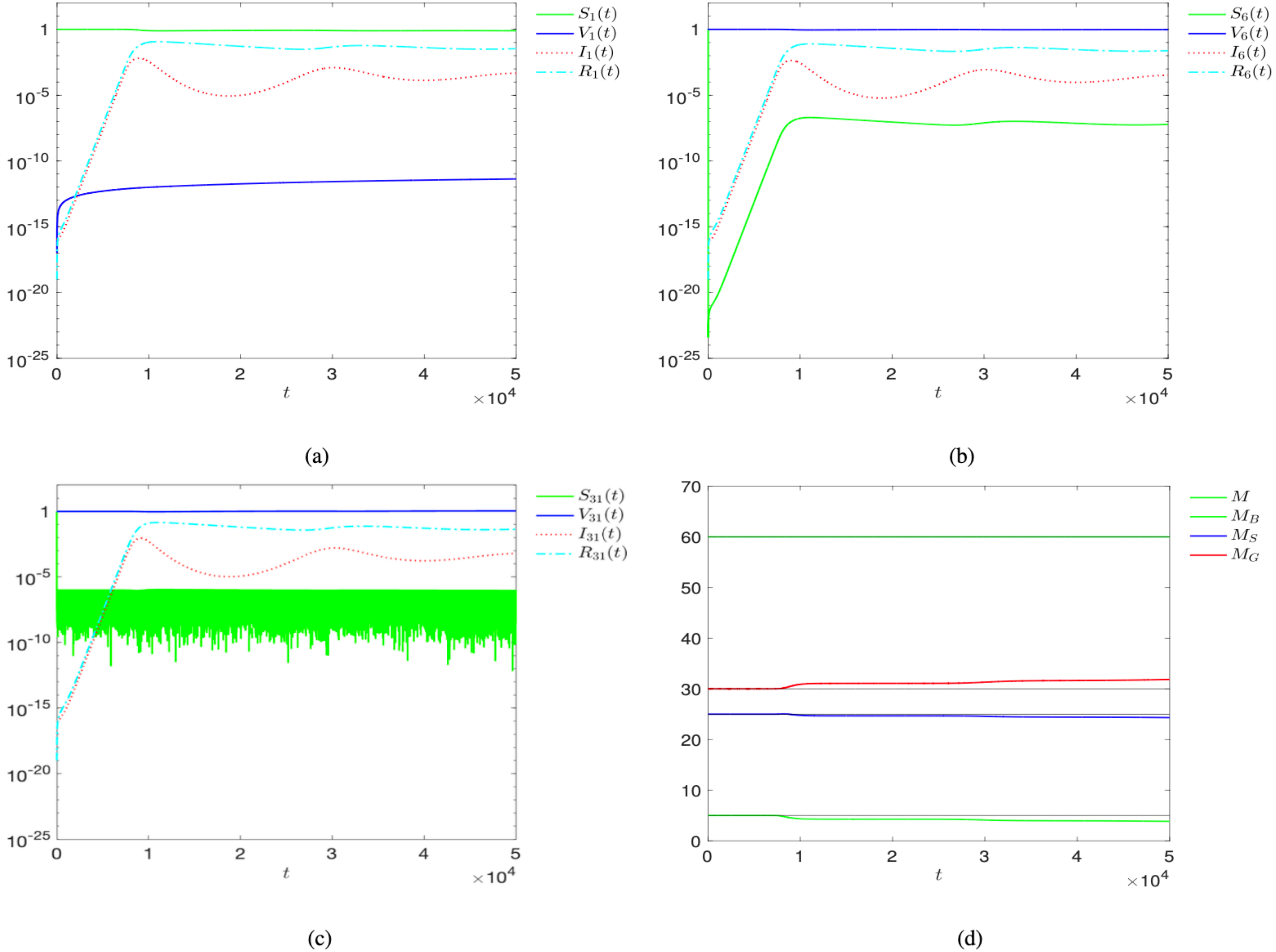}
\caption{Evolution of $S$, $V$, $I$, $R$, $M$, $M_B$, $M_S$ and $M_G$ in time for the configuration in panel (c) in Fig. \ref{fig_a2=0.01_thermalisation} for the BA scale-free network, where $\alpha=0.01$ in system \eqref{SVIR_coupled_system_ODEs}. Panel (a) shows $S_1$, $V_1$, $I_1$, $R_1$ in time of the first ``Bronze'' community. Panel (b) shows $S_6$, $V_6$, $I_6$, $R_6$ of the first ``Silver'' community in time and panel (c) shows $S_{31}$, $V_{31}$, $I_{31}$, $R_{31}$ of the first ``Gold'' community in time. Panel (d) shows the evolution of $M$, $M_B$, $M_S$ and $M_G$ in time. The black lines in panel (d) correspond to the number of ``Bronze'', ``Silver'' and ``Gold'' communities and to the total population, which are 5, 25, 30 and 60, respectively.}\label{fig_a2=0.01_thermalisation_BN}
\end{figure*}

\begin{figure*}
\centering
\includegraphics[height=17.1cm,width=0.99\textwidth]{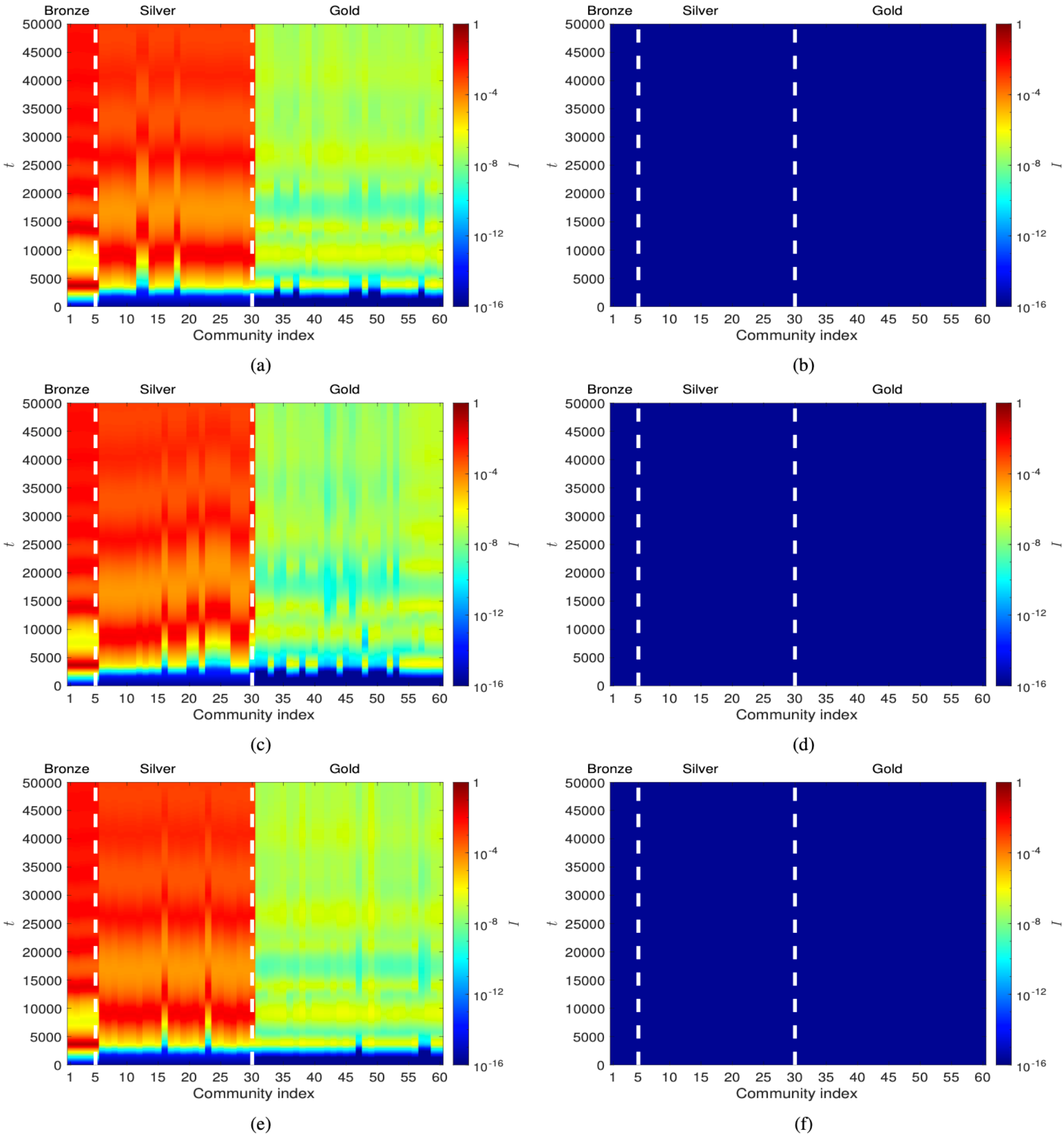}
 \caption{Spatiotemporal plots of infection levels for ER random, WS small-world and BA scale-free networks with network density 0.3 for $\alpha=10^{-8}$ and $\alpha=0.01$. Panels (a), (c), (e) show the spatiotemporal plot of infection levels for an ER random, WS small-world and BA scale-free network for $\alpha=10^{-8}$. Similarly, panels (b), (d), (f) the spatiotemporal plots for the same ER random, WS small-world and BA scale-free networks for $\alpha=0.01$. The scale of the colour bars is logarithmic to depict small variations in infections near 0 as infections $I$ take values in the interval $[0,1]$. The vertical white, dashed lines delineate the ranges of ``Bronze'', ``Silver'' and ``Gold'' communities.}\label{fig_sec_network_properties_and_the_spread_of_a_virus}
\end{figure*}

A similar analysis for the WS small-world network seen in Fig. \ref{fig_a2=0.01_thermalisation_WS}) and BA scale-free network in Fig. \ref{fig_a2=0.01_thermalisation_BN} shows that the infected populations approach orders of magnitude higher values (i.e., about $10^{-3}$ or $10^{-4}$) compared to those in the case of the ER random network in Fig. \ref{fig_a2=0.01_thermalisation_ERN}), which are smaller than $10^{-20}$. Figures \ref{fig_a2=0.01_thermalisation_WS}) and \ref{fig_a2=0.01_thermalisation_BN} also reveal that the vaccinated populations in the ``Bronze'' communities increase in time and saturate to higher values than initially, however that is not enough to keep at bay the infections, which see initially a dramatic increase before saturating to much higher values in the long term. Interestingly, panels (d) in Figs. \ref{fig_a2=0.01_thermalisation_WS}) and \ref{fig_a2=0.01_thermalisation_BN} for WS small-world and BA scale-free networks show that there is an increase over time in the number of ``Gold'' communities ($M_G$) and a decrease in the number of ``Bronze'' ($M_B$) and ``Silver`` ($M_S$) communities, keeping the total population constant in time at $N=60$ communities. This means that some of the ``Bronze'' and ``Silver'' communities turn into ``Gold'' communities, in contrast to the situation in ER random networks in Fig. \ref{fig_a2=0.01_thermalisation_ERN}, where that is not happening. The change in the number of ``Gold'' communities might be explained by a flow of populations moving from ``Bronze'' and ``Silver'' communities to ``Gold'' communities.

The results in Table I in the supplementary material confirm that the mean second smallest eigenvalue of the Laplacians of the adjacency matrices of the 100 ER random networks is bigger than the other types of networks used in Sec. \ref{subsec_stats_analysis}. This is also corroborated by the fact that the ER random, WS small-world and BA scale-free networks in Fig. \ref{fig_ERN_WSN_BN_60_nodes} have network densities 0.3, 0.07 and 0.0033, respectively. Next we have run simulations for ER random, BA scale-free and WS small-world networks with density 0.3 for coupling strengths $\alpha=10^{-8}$ in panels (a), (c), (e) in Fig. \ref{fig_sec_network_properties_and_the_spread_of_a_virus} and $\alpha=0.01$ in panels (b), (d), (f) in the same figure.

Our results in panels (a), (c), (e) show that for the small coupling strength $\alpha=10^{-8}$ and network density 0.3, the three types of communities reach different levels of infections, but most importantly they do not converge to the disease-free state. The situation changes drastically for coupling strength $\alpha=0.01$, as panels (b), (d), (f) show that all types of communities in all types of networks reach the disease-free state. These results are similar to those in Fig. 9(a) in the paper for the ER network which has the same network density as the networks in Fig. \ref{fig_sec_network_properties_and_the_spread_of_a_virus}. We conclude that it is the interplay between network structure (network density) and coupling strength that determine the dynamics and whether the system will approach the disease-free state.

Our analysis for $\alpha=0.01$ shows that the virus does not spread in densely-enough ER random, WS small-world and BA scale-free networks of connected communities as the system reaches the disease-free state. This is a striking result when compared to the spatiotemporal plots for the sparser WS small-world and BA scale-free networks for smaller coupling strength $\alpha=10^{-8}$, where the virus spreads to all communities in the long term, regardless of vaccination level and percentage and efficacy of the vaccination. We have also found that it takes more time for the virus to spread in a WS small-world network than in a BA scale-free network, presumably due to the presence of hubs (which act as mega-spreaders) in the latter network. Another aspect of our results is that the ``Gold'' communities are not able to slow-down the spread of the virus when they come in contact with the ``Bronze'' and ``Silver'' communities as in the long term, all types of communities in the WS small-world and BA scale-free networks become infected to similar high levels.

\section{Discussion and conclusions}\label{sec_discussion_conclusions}

In this paper, motivated by the ongoing COVID-19 pandemic, we sought to model mathematically the spread of a virus in interconnected communities (which can be for example countries) and explored different vaccination scenarios and their effect on the infection levels in the communities. Crucially, the infection levels depend on how the communities are interconnected.

We started assuming that all populations are initially susceptible and considered a susceptible-vaccinated-infected-re\-covered (SVIR) epidemiological model for each community, assuming that the vaccine efficacy wanes over time. We grouped the communities into unvaccinated (``Bronze''), moderately vaccinated (``Silver'') and very well vaccinated (``Gold'') and connected them via Erd\H{o}s-R\'enyi random, Bara\-b\'asi-Albert scale-free and Watts-Strogatz small-world networks via a diffusive linear coupling to emulate local spreading. We showed that when considering pair-wise interactions in ``Bronze''-``Gold" and ``Bronze''-``Silver" communities, the ``Bronze'' community is driving an increase in infections in the ``Silver'' and ``Gold'' communities, even though they are vaccinated. This shows the detrimental, one-way, effect of non-vaccinated communities to vaccinated ones when vaccine efficacy wanes over time. 

When considering the spread of a virus in larger-size networks characterised by hubs that act as mega-spreaders, i.e., in Barab\'asi-Albert networks, infections in the moderately vaccinated ``Silver'' and well-vaccinated ``Gold'' communities are much lower than those in the unvaccinated ``Bronze'' communities, leading to the latter being able to resist the spread of the virus for some time as the vaccine efficacy wanes over time. We find that the ``Gold'' communities are the best in keeping their infection levels at bay due to being the best vaccinated communities. However, our analysis shows that a small number of unvaccinated ``Bronze'' communities are enough to provoke an increase in infection levels in moderately and well-vaccinated communities. Even a large number of ``Gold'' communities is unable to reduce the levels of infections or halt the prevalence of the virus, driven by the non-vaccinated ``Bronze'' communities.

When studying the spread of a virus in Erd\H{o}s-R\'enyi, Bara\-b\'asi-Albert and Watts-Strogatz networks of strongly connected inhomogeneous communities with network densities 0.3, 0.07 and 0.033, respectively, the communities reach the disease-free state in Erd\H{o}s-R\'enyi networks, but not in the other two types of networks, as the infections spread to all communities in the long term. This might be possible because of long-range connections in small-world networks and hubs in scale-free networks, which act as mega-spreaders. However, when all types of networks have the same density, then the dynamics in all three types of networks reach the disease-free state.

Based on our results, we conclude that if a community is not vaccinated periodically in time, due to the vaccination efficacy waning over time, it will become infected in the long term when connected to other infected populations. This is alarming and revealing at the same time of what is required of countries at a global scale to defend their populations against the spread of infectious diseases such as COVID-19. Implementing vaccination programs globally and vaccinating as many individuals as possible is the way to safeguard populations from the spread of a virus.

Our study contributes to the understanding of the spread of viruses by proposing a generic mathematical model for the evolution of pandemics in networks of interconnected communities. The SVIR model used in this paper has been chosen to showcase the role of vaccination, since it is an important factor that can help reduce deaths due to the virus. Our work brings together two areas of well-established research investigations: (a) nonlinear models of epidemiology and (b) models of complex network dynamics, hence its novelty lies in the integration of the two areas.

Our investigations show how a simple but generic mathematical model such as model \eqref{SVIR_coupled_system_ODEs} can demonstrate the empirically verified and well understood in terms of scientific and medical reasoning, fact that ``nobody is safe until everybody is safe''. Our goal is not to propose a detailed and complicated model with a plethora of variables and parameters. We want to highlight the generic properties of the complex evolution of pandemics in an interconnected world of communities. Adopting a modular approach whereby model \eqref{svir_uncoupled} can be replaced by other relevant models and by sharing our numerical code on GitHub \cite{githubSVIRNet2020}, we hope our work can be used to help understand the spread of viruses in interconnected communities, at large.

Our work can be extended at the level of (a) modules: variants of the main SVIR model can include dynamical aspects such as time delays, state-dependent control, incorporation of mutations of the infecting species, etc. that have already been considered in the literature but also new ones, (b) graph/network structure, as different networks, even realistic ones, can be used to couple the communities through the Laplacian, (c) coupling as it can incorporate more realistic, weighted sums, or state-depend control-types of coupling, reflecting choices of strategies for the containment of the spread of a virus at a local level, mesoscopically or globally.

Model \eqref{SVIR_coupled_system_ODEs} for pandemics in networks of communities, proposed in this work, highlights one of the simplest and generic ways and the role of vaccination as the utmost key factors of containing a pandemic.

Finally, it would be interesting to study in a future publication, the analytical solutions in the neighbourhood of the equilibrium points of systems \eqref{eq_classic_SIR_Model}, \eqref{eq_original_single_SVIR_Model}, \eqref{svir_uncoupled} by linearising them and using matrix or power-series methods. A relevant approach is discussed in \cite{Ogunmiloro2019NUMERICALAS} and could be used to derive analytical conditions for the convergence of the solutions to the equilibrium points and the disease-free states.

\section*{Supplementary material}

In the supplementary material, we present additional results on the spread of a virus in different types of networks, i.e., in Erd\H{o}s-R\'enyi (ER) random \cite{Erdosetal1959}, Watts-Strogatz (WS) small-world \cite{Wattsetal1998} and Barab\'asi-Albert (BA) scale-free \cite{RevModPhys.74.47} networks and on network properties that are not included in the paper. In Sec. II, we present results for communities sorted based on the node-degrees in a BA scale-free network, in Sec. III, results for the case where only susceptible populations can travel through a network and finally, in Sec. IV, we discuss some of the spectral properties of networks used in the paper.

\section*{Data availability}
Data available in article or supplementary material.

\section*{Author declaration}
The authors have no conflicts to disclose.



\nocite{*}
\bibliographystyle{plainurl}

\begin{thebibliography}{10}

\bibitem{RevModPhys.74.47}
R.~Albert and A-L. Barab\'asi.
\newblock Statistical mechanics of complex networks.
\newblock {\em Rev. Mod. Phys.}, 74:47--97, 2002.

\bibitem{alcaraz2020}
G.~G. Alcaraz and C.~Vargas-De-Le{\'o}n.
\newblock Modeling control strategies for influenza {A H1N1} epidemics: {SIR}
  models.
\newblock {\em Revista Mexicana de F{\'\i}sica}, 58(1):37--43, 2012.

\bibitem{Almeshaletal2020}
A.~M. Almeshal, A.~I. Almazrouee, M.~R. Alenizi, and S.~N. Alhajeri.
\newblock Forecasting the spread of {COVID-19} in {Kuwait} using compartmental
  and logistic regression models.
\newblock {\em Appl. Sci.}, 10:3402, 2020.

\bibitem{amaro2021global}
J.~E. Amaro, J.~Dudouet, and J.~N. Orce.
\newblock Global analysis of the {COVID-19} pandemic using simple
  epidemiological models.
\newblock {\em Applied Mathematical Modelling}, 90:995, 2021.

\bibitem{Anas2020}
C.~Anastassopoulou, L.~Russo, A.~Tsakris, and C.~Siettos.
\newblock Data-based analysis, modelling and forecasting of the {COVID-19}
  outbreak.
\newblock {\em PLOS ONE}, 15(3):1--21, 03 2020.

\bibitem{Anderson1991}
R.~M. Anderson.
\newblock Discussion: {The Kermack-McKendrick} epidemic threshold theorem.
\newblock {\em Bulletin of Mathematical Biology}, 53:1, 1991.

\bibitem{arino2003}
J.~Arino, C.~C. McCluskey, and P.~van~den Driessche.
\newblock Global results for an epidemic model with vaccination that exhibits
  backward bifurcation.
\newblock {\em SIAM Journal on Applied Mathematics}, 64(1):260--276, 2003.

\bibitem{bagal2020estimating}
D.~K. Bagal, A.~Rath, A.~Barua, and D.~Patnaik.
\newblock Estimating the parameters of susceptible-infected-recovered model of
  {COVID-19} cases in {India} during lockdown periods.
\newblock {\em Chaos, Solitons \& Fractals}, 140:110154, 2020.

\bibitem{Bar-Zeev2020}
N.~Bar-Zeev and T.~Inglesby.
\newblock {COVID-19} vaccines: {Early} success and remaining challenges.
\newblock {\em The Lancet}, 396:868, 2020.

\bibitem{basu2020going}
S.~Basu and R.~H. Campbell.
\newblock Going by the numbers: {Learning} and modeling {COVID-19} disease
  dynamics.
\newblock {\em Chaos, Solitons \& Fractals}, 138:110140, 2020.

\bibitem{boccaletti2020modeling}
S.~Boccaletti, W.~Ditto, G.~Mindlin, and A.~Atangana.
\newblock Modeling and forecasting of epidemic spreading: {The} case of
  {COVID-19} and beyond.
\newblock {\em Chaos, Solitons \& Fractals}, 135:109794, 2020.

\bibitem{Braueretal2019}
F.~Brauer, C.~Castillo-Chavez, and Z.~Feng.
\newblock {\em Mathematical Models in Epidemiology}.
\newblock Springer, 2019.

\bibitem{cooper2020dynamic}
I.~Cooper, A.~Mondal, and C.~G. Antonopoulos.
\newblock Dynamic tracking with model-based forecasting for the spread of the
  {COVID-19} pandemic.
\newblock {\em Chaos, Solitons \& Fractals}, 139:110298, 2020.

\bibitem{cooper2020sir}
I.~Cooper, A.~Mondal, and C.~G. Antonopoulos.
\newblock A {SIR} model assumption for the spread of {COVID-19} in different
  communities.
\newblock {\em Chaos, Solitons \& Fractals}, 139:110057, 2020.

\bibitem{VARGASDELEON20111106}
V.~L. Cruz.
\newblock On the global stability of {SIS, SIR and SIRS} epidemic models with
  standard incidence.
\newblock {\em Chaos, Solitons \& Fractals}, 44(12):1106--1110, 2011.

\bibitem{das2021covid}
A.~Das, A.~Dhar, S.~Goyal, A.~Kundu, and S.~Pandey.
\newblock {COVID-19: Analytic} results for a modified {SEIR} model and
  comparison of different intervention strategies.
\newblock {\em Chaos, Solitons \& Fractals}, 144:110595, 2021.

\bibitem{dashtbali2021compartmental}
M.~Dashtbali and M.~Mirzaie.
\newblock A compartmental model that predicts the effect of social distancing
  and vaccination on controlling {COVID-19}.
\newblock {\em Scientific Reports}, 11(1):1, 2021.

\bibitem{dehning2020inferring}
J.~Dehning, J.~Zierenberg, F.~P. Spitzner, M.~Wibral, J.~P. Neto, M.~Wilczek,
  and V.~Priesemann.
\newblock Inferring change points in the spread of {COVID-19} reveals the
  effectiveness of interventions.
\newblock {\em Science}, 369(6500), 2020.

\bibitem{Dormandetal1980}
J.~R. Dormand and P.~J. Prince.
\newblock A family of embedded {Runge-Kutta} formulae.
\newblock {\em J. Comp. Appl. Math.}, 6:19--26, 1980.

\bibitem{Erdosetal1959}
P.~Erd\H{o}s and A.~R\'enyi.
\newblock On random graphs.
\newblock {\em Publicationes Mathematicae}, 6:290--297, 1959.

\bibitem{fanelli2020analysis}
D.~Fanelli and F.~Piazza.
\newblock Analysis and forecast of {COVID-19} spreading in {China, Italy and
  France}.
\newblock {\em Chaos, Solitons \& Fractals}, 134:109761, 2020.

\bibitem{Gakkhar2008}
S.~Gakkhar and K.~Negi.
\newblock Pulse vaccination in {SIRS} epidemic model with non-monotonic
  incidence rate.
\newblock {\em Chaos, Solitons \& Fractals}, 35(3):626--638, 2008.

\bibitem{Harkoetal2014}
T.~Harko, F.~S.~N. Lobo, and M.~K. Mak.
\newblock Exact analytical solutions of the {Susceptible-Infected-Recovered
  (SIR)} epidemic model and of the {SIR} model with equal death and birth
  rates.
\newblock {\em Appl. Math. Comput.}, 236:184, 2014.

\bibitem{Hethcote1989}
H.~W. Hethcote.
\newblock {\em Three Basic Epidemiological Models}.
\newblock Springer Berlin Heidelberg, Berlin, Heidelberg, 1989.

\bibitem{Hethcote2000}
H.~W. Hethcote.
\newblock The mathematics of infectious diseases.
\newblock {\em SIAM Review}, 42:599, 2000.

\bibitem{Hethcote2008}
H.~W. Hethcote.
\newblock The basic epidemiology models: {Models}, expressions for {$R_0$},
  parameter estimation, and applications.
\newblock {\em Mathematical Understanding of Infectious Disease Dynamics},
  16:1, 2008.

\bibitem{Hou2020}
C.~Hou, J.~Chen, Y.~Zhou, L.~Hua, J.~Yuan, S.~He, Y.~Guo, S.~Zhang, Q.~Jia,
  C.~Zhao, J.~Zhang, G.~Xu, and E.~Jia.
\newblock The effectiveness of quarantine of {Wuhan} city against the {Corona
  Virus Disease 2019 (COVID-19): A} well-mixed {SEIR} model analysis.
\newblock {\em Journal of Medical Virology}, 92(7):841--848, 2020.

\bibitem{Kermacketal2020}
W.~O. Kermack and A.~Mckendrick.
\newblock A contribution to the mathematical theory of epidemics.
\newblock {\em Proc. R. Soc. A Math. Phys. Eng. Sci.}, 115:700, 2020.

\bibitem{Khoshnawetal2020}
S.~H.~A. Khoshnaw, M.~Shahzad, M.~Ali, and Sultan F.
\newblock Quantitative and qualitative analysis of the {COVID-19} pandemic
  model.
\newblock {\em Chaos Solitons \& Fract.}, 20:109932, 2020.

\bibitem{kobayashi2020predicting}
G.~Kobayashi, S.~Sugasawa, H.~Tamae, and T.~Ozu.
\newblock Predicting intervention effect for {COVID-19} in {Japan: State} space
  modeling approach.
\newblock {\em BioScience Trends}, 2020.

\bibitem{Kraemer2020}
M.~U.~G. Kraemer, C.~H. Yang, B.~Gutierrez, C.~H. Wu, B.~Klein, D.~M. Pigott,
  Open COVID-19 Data~Working Group, L.~du~Plessis, N.~R. Faria, R.~Li, W.~P.
  Hanage, J.~S. Brownstein, M.~Layan, A.~Vespignani, H.~Tian, C.~Dye, O.~G.
  Pybus, and S.~V. Scarpino.
\newblock The effect of human mobility and control measures on the {COVID-19}
  epidemic in {China}.
\newblock {\em Science}, 368(6490):493, 2020.

\bibitem{Kupferschmidt1375}
K.~Kupferschmidt and M.~Wadman.
\newblock Delta variant triggers new phase in the pandemic.
\newblock {\em Science}, 372(6549):1375--1376, 2021.

\bibitem{liao2020tw}
Z.~Liao, P.~Lan, Z.~Liao, Y.~Zhang, and S.~Liu.
\newblock {TW-SIR: Time-window} based {SIR} for {COVID-19} forecasts.
\newblock {\em Scientific Reports}, 10(1):1, 2020.

\bibitem{liu2017}
L.~Liu, X.~Luo, and L.~Chang.
\newblock Vaccination strategies of an {SIR} pair approximation model with
  demographics on complex networks.
\newblock {\em Chaos, Solitons \& Fractals}, 104:282--290, 2017.

\bibitem{LIU20081}
X.~Liu, Y.~Takeuchi, and S.~Iwami.
\newblock {SVIR} epidemic models with vaccination strategies.
\newblock {\em Journal of Theoretical Biology}, 253(1):1--11, 2008.

\bibitem{liu2021predicting}
Z.~Liu, P.~Magal, and G.~Webb.
\newblock Predicting the number of reported and unreported cases for the
  {COVID-19} epidemics in {China, South Korea, Italy, France, Germany and
  United Kingdom}.
\newblock {\em Journal of Theoretical Biology}, 509:110501, 2021.

\bibitem{LOPEZ2021103746}
L.~L\'opez and X.~Rod\'o.
\newblock A modified {SEIR} model to predict the {COVID-19} outbreak in {Spain
  and Italy: Simulating} control scenarios and multi-scale epidemics.
\newblock {\em Results in Physics}, 21:103746, 2021.

\bibitem{machado2020nonlinear}
J.~A.~T. Machado and J.~Ma.
\newblock Nonlinear dynamics of {COVID-19} pandemic: {Modeling}, control, and
  future perspectives.
\newblock {\em Nonlinear Dynamics}, 101(3):1525, 2020.

\bibitem{MALAVIKA202126}
B.~Malavika, S.~Marimuthu, M.~Joy, A.~Nadaraja, E.~S. Asirvathamb, and
  L.~Jeyaseelanc.
\newblock Forecasting {COVID-19} epidemic in {India} and high incidence states
  using {SIR} and logistic growth models.
\newblock {\em Clinical Epidemiology and Global Health}, 9:26, 2021.

\bibitem{Miaoetal2017}
A.~Miao, J.~Zhang, T.~Zhang, and B.~G. Pradeep.
\newblock Threshold dynamics of a stochastic model with vertical transmission
  and vaccination.
\newblock {\em Comput. Math. Methods Med.}, 2017(3), 2017.

\bibitem{mishra2021mathematical}
B.~K. Mishra, A.~K. Keshri, D.~K. Saini, S.~Ayesha, B.~K. Mishra, and Y.~S.
  Rao.
\newblock Mathematical model, forecast and analysis on the spread of
  {COVID-19}.
\newblock {\em Chaos, Solitons \& Fractals}, 147:110995, 2021.

\bibitem{munoz2021sir}
G.~A. Mu{\~n}oz-Fern{\'a}ndez, J.~M. Seoane, and J.~B. Seoane-Sep{\'u}lveda.
\newblock A {SIR-type} model describing the successive waves of {COVID-19}.
\newblock {\em Chaos, Solitons \& Fractals}, 144:110682, 2021.

\bibitem{Ndairou2020}
F.~Nda\"irou, I.~Area, J.~J. Nieto, and D.~F.~M. Torres.
\newblock Mathematical modeling of {COVID-19} transmission dynamics with a case
  study of {Wuhan}.
\newblock {\em Chaos, Solitons \& Fractals}, 135:109846, 2020.

\bibitem{Newman2010}
M.~Newman.
\newblock {\em Networks: {An} introduction}.
\newblock Oxford University Press, 2010.

\bibitem{odagaki2020analysis}
T.~Odagaki.
\newblock Analysis of the outbreak of {COVID-19 in Japan} by {SIQR} model.
\newblock {\em Infectious Disease Modelling}, 5:691, 2020.

\bibitem{odagaki2021classification}
T.~Odagaki and R.~Suda.
\newblock Classification of the infection status of {COVID-19} in 190
  countries.
\newblock {\em medRxiv}, pages 2020--12, 2021.

\bibitem{Ogunmiloro2019NUMERICALAS}
M.~O. Oluwatayo, O.~A. Fatima, and Kareem~H. A.
\newblock Numerical and stability analysis of the transmission dynamics of
  {SVIR} epidemic model with standard incidence rate.
\newblock volume~4, pages 349--361, 2019.

\bibitem{pai2020investigating}
C.~Pai, A.~Bhaskar, and V.~Rawoot.
\newblock Investigating the dynamics of {COVID-19} pandemic in {India} under
  lockdown.
\newblock {\em Chaos, Solitons \& Fractals}, 138:109988, 2020.

\bibitem{Peaketal2020}
C.~M. Peak, R.~Kahn, Y.~H. Grad, L.~M. Childs, R.~Li, M.~Lipsitch, and D.~C.~O.
  Buckee.
\newblock Individual quarantine versus active monitoring of contacts for the
  mitigation of {COVID-19}: {A} modelling study.
\newblock {\em Lancet Infect Dis.}, 20:1025--1033, 2020.

\bibitem{peng2013}
X-L. Peng, X-J. Xu, X.~Fu, and T.~Zhou.
\newblock Vaccination intervention on epidemic dynamics in networks.
\newblock {\em Physical Review E}, 87(2):022813, 2013.

\bibitem{peng2016}
X-L. Peng, X-J. Xu, M.~Small, X.~Fu, and Z.~Jin.
\newblock Prevention of infectious diseases by public vaccination and
  individual protection.
\newblock {\em Journal of mathematical biology}, 73(6):1561--1594, 2016.

\bibitem{peng2019}
X-L. Peng, Z-Q. Zhang, J.~Yang, and Z.~Jin.
\newblock An {SIS} epidemic model with vaccination in a dynamical contact
  network of mobile individuals with heterogeneous spatial constraints.
\newblock {\em Communications in Nonlinear Science and Numerical Simulation},
  73:52--73, 2019.

\bibitem{postnikov2020estimation}
E.~B. Postnikov.
\newblock Estimation of {COVID-19} dynamics ``on a back-of-envelope'': {Does}
  the simplest {SIR} model provide quantitative parameters and predictions?
\newblock {\em Chaos, Solitons \& Fractals}, 135:109841, 2020.

\bibitem{rafiq2020evaluation}
D.~Rafiq, S.~A. Suhail, and M.~A. Bazaz.
\newblock Evaluation and prediction of {COVID-19} in {India: A} case study of
  worst hit states.
\newblock {\em Chaos, Solitons \& Fractals}, 139:110014, 2020.

\bibitem{ranjan2020predictions}
R.~Ranjan.
\newblock Predictions for {COVID-19} outbreak in {India} using epidemiological
  models.
\newblock {\em MedRxiv}, 2020.

\bibitem{samui2020mathematical}
P.~Samui, J.~Mondal, and S.~Khajanchi.
\newblock A mathematical model for {COVID-19} transmission dynamics with a case
  study of {India}.
\newblock {\em Chaos, Solitons \& Fractals}, 140:110173, 2020.

\bibitem{sardar2020assessment}
T.~Sardar, S.~S. Nadim, S.~Rana, and J.~Chattopadhyay.
\newblock Assessment of lockdown effect in some states and overall {India:
  Predictive} mathematical study on {COVID-19} outbreak.
\newblock {\em Chaos, Solitons \& Fractals}, 139:110078, 2020.

\bibitem{sarkar2020modeling}
K.~Sarkar, S.~Khajanchi, and J.~J. Nieto.
\newblock Modeling and forecasting the {COVID-19} pandemic in {India}.
\newblock {\em Chaos, Solitons \& Fractals}, 139:110049, 2020.

\bibitem{Sayama2015}
H.~Sayama.
\newblock {\em Introduction to the Modeling and Analysis of Complex Systems}.
\newblock Open SUNY Textbooks, 2015.

\bibitem{scarpino2019predictability}
S.~V. Scarpino and G.~Petri.
\newblock On the predictability of infectious disease outbreaks.
\newblock {\em Nature Communications}, 10(1):1, 2019.

\bibitem{scherer2002}
A.~Scherer and A.~McLean.
\newblock Mathematical models of vaccination.
\newblock {\em British Medical Bulletin}, 62(1):187--199, 2002.

\bibitem{10.3389/fmed.2020.00171}
L.~Su, N.~Hong, X.~Zhou, J.~He, Y.~Ma, H.~Jiang, L.~Han, F.~Chang, G.~Shan,
  W.~Zhu, and Y.~Long.
\newblock Evaluation of the secondary transmission pattern and epidemic
  prediction of {COVID-19} in the four metropolitan areas of {China}.
\newblock {\em Frontiers in Medicine}, 7:171, 2020.

\bibitem{tang2020estimation}
B.~Tang, X.~Wang, Q.~Li, N.~L. Bragazzi, S.~Tang, Y.~Xiao, and J.~Wu.
\newblock Estimation of the transmission risk of the {2019-nCoV} and its
  implication for public health interventions.
\newblock {\em Journal of clinical medicine}, 9(2):462, 2020.

\bibitem{Tornatoreetal2014}
E.~Tornatore, P.~Vetro, and S.~M. Buccellato.
\newblock {SIVR} epidemic model with stochastic perturbation.
\newblock {\em Neural Comput. \& Applic.}, 24:309--315, 2014.

\bibitem{Tuiteetal2020}
A.~R. Tuite, D.~N. Fisman, and A.~L. Greer.
\newblock Mathematical modelling of {COVID-19} transmission and mitigation
  strategies in the population of {Ontario, Canada}.
\newblock {\em CMAJ}, 192:E497--E505, 2020.

\bibitem{tyagi2021mathematical}
S.~Tyagi, S.~C. Martha, S.~Abbas, and A.~Debbouch.
\newblock Mathematical modeling and analysis for controlling the spread of
  infectious diseases.
\newblock {\em Chaos, Solitons \& Fractals}, 144:110707, 2021.

\bibitem{COVID19-India_website}
\url{covid19india.org,github.com/covid19india/api}.
\newblock Covid-19 {India}.
\newblock {\em COVID-19 India}, 2021.

\bibitem{Who2020}
\url{https://covid19.who.int/}.
\newblock {WHO} coronavirus {(COVID-19)} dashboard.
\newblock {\em World Health Organization}, 2020.

\bibitem{Who2021b}
\url{https://covid19.who.int}.
\newblock {WHO} coronavirus {(COVID-19)} dashboard.
\newblock {\em World Health Organization}, 2021.

\bibitem{githubSVIRNet2020}
\url{https://github.com/Alcamis/SVIR_Net}.
\newblock {SVIR N}et.
\newblock {\em GitHub}, 2021.

\bibitem{Who2021a}
\url{https://www.euro.who.int/en/health-topics/health-emergencies/coronavirus-covid-19/covid-19-vaccines-and-vaccination}.
\newblock {COVID-19} vaccines and vaccination.
\newblock {\em World Health Organization}, 2021.

\bibitem{ONS2021}
\url{https://www.ons.gov.uk/peoplepopulationandcommunity/healthandsocialcare/conditionsanddiseases/articles/coronaviruscovid19/latestinsights}.
\newblock Coronavirus {(COVID-19)} latest insights.
\newblock {\em Office for National Statistics UK}, 2021.

\bibitem{Who2019}
\url{https://www.who.int/emergencies/diseases/novel-coronavirus-2019}.
\newblock Coronavirus disease {(COVID-19)} outbreak.
\newblock {\em World Health Organization}, 2019.

\bibitem{Worldometer_website}
\url{https://www.worldometers.info/coronavirus/}.
\newblock Coronavirus worldometer website.
\newblock {\em Worldometer}, 2021.

\bibitem{vaishnav2020assessment}
V.~Vaishnav and J.~Vajpai.
\newblock Assessment of impact of relaxation in lockdown and forecast of
  preparation for combating {COVID-19} pandemic in {India} using group method
  of data handling.
\newblock {\em Chaos, Solitons \& Fractals}, 140:110191, 2020.

\bibitem{wang2019}
X.~Wang, H.~Peng, B.~Shi, D.~Jiang, S.~Zhang, and Bi. Chen.
\newblock Optimal vaccination strategy of a constrained time-varying {SEIR}
  epidemic model.
\newblock {\em Communications in Nonlinear Science and Numerical Simulation},
  67:37--48, 2019.

\bibitem{Wattsetal1998}
D.~Watts and S.~Strogatz.
\newblock Collective dynamics of ``small-world'' networks.
\newblock {\em Nature}, 393:440--442, 1998.

\bibitem{Weiss2013}
H.~H. Weiss.
\newblock The mathematics of infectious diseases.
\newblock {\em SIAM Review}, 42:599, 2000.

\bibitem{weiss2013SIR}
H.~H. Weiss.
\newblock The {SIR} model and the foundations of public health.
\newblock {\em MATerials MATem\`atics}, pages 0001--17, 2013.

\bibitem{wu2020new}
F.~Wu, S.~Zhaio, B.~Yu, Y-M. Chen, W.~Wang, Z-G. Song, Y.~Hu, Z-W. Tao, J-H.
  Tian, P.~Yuan-Yuan, L.~Yuan, Y-L. Zhang, F-H. Dai, Y.~Liu, Q-M. Wang, J-J.
  Zheng, L.~Xu, E.~C. Holmes, and Y-Z. Zhang.
\newblock A new coronavirus associated with human respiratory disease in
  {China}.
\newblock {\em Nature}, 579(7798):265, 2020.

\bibitem{wu2020nowcasting}
J.~T. Wu, K.~Leung, and G.~M. Leung.
\newblock Nowcasting and forecasting the potential domestic and international
  spread of the {2019-nCoV} outbreak originating in {Wuhan, China}: {A}
  modelling study.
\newblock {\em The Lancet}, 395(10225):689, 2020.

\end{thebibliography}

\end{document}